\documentclass[10pt,journal,twoside,final]{IEEEtran}
\usepackage[T1]{fontenc}
\usepackage{ifpdf}
\usepackage{color}
\usepackage{psfrag}
\usepackage[dvips]{graphicx}
\usepackage{tikz}
\usepackage{pgfplots}
\pgfplotsset{compat=newest}
\usepackage{enumerate}
\usepackage{srcltx}
\usepackage[noadjust]{cite}
\usepackage[cmex10]{amsmath}
\usepackage{amssymb}
\usepackage{ntheorem}
\usepackage{dsfont}    
\usepackage{mathtools}
\usepackage{algorithmic}
\usepackage{array}
\newcommand{\ve}[1]{{\boldsymbol{#1}}}
\newcommand{\ma}[1]{{\boldsymbol{#1}}}

\newcommand{\Prob}{\mathop{{\sf P}}\nolimits}

\newcommand{\e}{\mathop{\mathrm{e}}\nolimits}

\newcommand{\argmin}{\operatornamewithlimits{argmin}}

\newcommand{\Nomo}{\mathop{\mathsf{N}}\nolimits}
\newcommand{\C}{\mathop{\mathsf{C}}\nolimits}
\newcommand{\F}{\mathop{\mathsf{F}}\nolimits}
\newcommand{\Vol}{\mathop{\mathrm{Vol}}\nolimits}

\theoremstyle{plain}
\theoremheaderfont{\itshape}\theorembodyfont{\itshape}
\theoremseparator{.}
\newtheorem{theorem}{Theorem}

\newtheorem{lemma}{Lemma}

\theoremheaderfont{\itshape}\theorembodyfont{\upshape}
\theoremseparator{.}
\newtheorem{definition}{Definition}
\newtheorem{remark}{Remark}
\newtheorem{example}{Example}

\usepackage[tight,footnotesize]{subfigure}
\usepackage{todonotes}
\begin{document}
\title{Nomographic Functions: Efficient Computation in Clustered Gaussian Sensor Networks}%%
\author{Mario~Goldenbaum,~\IEEEmembership{Student~Member,~IEEE,} Holger~Boche,~\IEEEmembership{Fellow,~IEEE}, and\\ S\l awomir~Sta\'{n}czak,~\IEEEmembership{Senior Member,~IEEE}%
\thanks{\copyright 2014 IEEE. Personal use of this material is permitted. Permission from IEEE must be obtained for all other uses, in any current or future media, including reprinting/republishing this material for advertising or promotional purposes, creating new collective works, for resale or redistribution to servers or lists, or reuse of any copyrighted component of this work in other works.}
\thanks{The work of M. Goldenbaum and S. Sta\'{n}czak was supported in part by the German Federal Ministry for Economic Affairs and Energy (BMWi) under grant 01MA13008B and the German Research Foundation (DFG) under grant STA 864/3-2. The work of H. Boche was supported in part by the DFG under grant Bo 1734/20-1. Parts of the material in this paper were presented at the 2013 IEEE Int. Conf. Acoustics, Speech and Signal Processing (ICASSP) \cite{Goldenbaum:Boche:Stanczak:13a}.}
\thanks{M.\,Goldenbaum is with the Communications and Information Theory Chair, Technische Universität Berlin, Berlin, Germany, and with the Lehrstuhl für Theoretische Informationstechnik, Technische Universität München, Munich, Germany (e-mail: mario.goldenbaum@tu-berlin.de).}%%
\thanks{H.\,Boche is with the Lehrstuhl für Theoretische Informationstechnik, Technische Universität München, Munich, Germany (e-mail: boche@tum.de).}%
\thanks{S.\,Sta\'{n}czak is with the Fraunhofer Institute for Telecommunications, Heinrich Hertz Institute, Berlin, Germany, and with the Communications and Information Theory Chair, Technische Universität Berlin, Berlin, Germany (e-mail: slawomir.stanczak@hhi.fraunhofer.de).}}%%
%
%\IEEEpubid{}%
%\markboth{IEEE TRANSACTIONS ON WIRELESS COMMUNICATIONS}{GOLDENBAUM \MakeLowercase{\textit{et al.}}: Nomographic Functions: Reliable Computation in Clustered Gaussian Sensor Networks}% 
\IEEEaftertitletext{\vspace{-1\baselineskip}}%%
\maketitle %%
%
%
%
%
%%%%%%%%%%%%%%%%%%%%%%%%%%%%%%%%%%%%%%%%%%%%%%%%%%%%%%%%%%%%%%%%%%%%%%%%%%%%%%%%%%%%%%%%%%%%%%%%%%%%%%%%%%%%%%%%%%%%%%%%%%%
%%%%%%%%%%%%%%%%%%%%%%%%%%%%%%%%%%%%%%%%%%%%%%%%%%%%%%%%%%%%%%%%%%%%%%%%%%%%%%%%%%%%%%%%%%%%%%%%%%%%%%%%%%%%%%%%%%%%%%%%%%%
%	Abstract
%%%%%%%%%%%%%%%%%%%%%%%%%%%%%%%%%%%%%%%%%%%%%%%%%%%%%%%%%%%%%%%%%%%%%%%%%%%%%%%%%%%%%%%%%%%%%%%%%%%%%%%%%%%%%%%%%%%%%%%%%%%
%%%%%%%%%%%%%%%%%%%%%%%%%%%%%%%%%%%%%%%%%%%%%%%%%%%%%%%%%%%%%%%%%%%%%%%%%%%%%%%%%%%%%%%%%%%%%%%%%%%%%%%%%%%%%%%%%%%%%%%%%%%
%
\begin{abstract}%
	In this paper, a clustered wireless sensor network is considered that is modeled as a set of coupled Gaussian multiple-access channels. The objective of the network is not to reconstruct individual sensor readings at designated fusion centers but rather to reliably compute some functions thereof. Our particular attention is on real-valued functions that can be represented as a post-processed sum of pre-processed sensor readings. Such functions are called nomographic functions and their special structure permits the utilization of the interference property of the Gaussian multiple-access channel to reliably compute many linear and nonlinear functions at significantly higher rates than those achievable with standard schemes that combat interference. Motivated by this observation, a computation scheme is proposed that combines a suitable data pre- and post-processing strategy with a nested lattice code designed to protect the sum of pre-processed sensor readings against the channel noise. After analyzing its computation rate performance, it is shown that at the cost of a reduced rate, the scheme can be extended to compute every continuous function of the sensor readings in a finite succession of steps, where in each step a different nomographic function is computed. This demonstrates the fundamental role of nomographic representations.%
\end{abstract}%
%
%
%
%%%%%%%%%%%%%%%%%%%%%%%%%%%%%%%%%%%%%%%%%%%%%%%%%%%%%%%%%%%%%%%%%%%%%%%%%%%%%%%%%%%%%%%%%%%%%%%%%%%%%%%%%%%%%%%%%%%%%%%%%%%
%%%%%%%%%%%%%%%%%%%%%%%%%%%%%%%%%%%%%%%%%%%%%%%%%%%%%%%%%%%%%%%%%%%%%%%%%%%%%%%%%%%%%%%%%%%%%%%%%%%%%%%%%%%%%%%%%%%%%%%%%%%
%	Introduction
%%%%%%%%%%%%%%%%%%%%%%%%%%%%%%%%%%%%%%%%%%%%%%%%%%%%%%%%%%%%%%%%%%%%%%%%%%%%%%%%%%%%%%%%%%%%%%%%%%%%%%%%%%%%%%%%%%%%%%%%%%%
%%%%%%%%%%%%%%%%%%%%%%%%%%%%%%%%%%%%%%%%%%%%%%%%%%%%%%%%%%%%%%%%%%%%%%%%%%%%%%%%%%%%%%%%%%%%%%%%%%%%%%%%%%%%%%%%%%%%%%%%%%%
%
\section{Introduction} \label{sec:intro}
\IEEEPARstart{M}{any} wireless sensor network applications require a reliable computation of application-dependent functions of the sensor readings at one or multiple fusion centers (e.g., arithmetic mean, maximum value) \cite{Giridhar:Kumar:05}. To solve such a distributed computation problem, the access of nodes to the common channel is usually coordinated so that the fusion centers can reconstruct individual sensor readings in an interference-free manner, followed by a subsequent computation of the function-values of interest. In what follows, we call such computation strategies (i.e., strategies that combat interference to recover all the associated sensor readings at the receiver side) \emph{separation-based} approaches as they strictly separate the wireless communication from the process of computation.

In the seminal paper \cite{Nazer:Gastpar:07b}, it is shown that this approach can be highly inefficient when the function to be computed at the fusion center is \emph{linear}. More precisely, it is shown that the interference caused by concurrent transmissions can be harnessed to compute function values at significantly higher rates than those achievable with separation-based strategies.

The problem of exploiting interference for efficiently computing \emph{nonlinear} functions of the sensor readings is addressed in \cite{Goldenbaum:Stanczak:13a}. The main idea of the scheme proposed in \cite{Goldenbaum:Stanczak:13a} is to apply an appropriate pre-processing function to each real-valued sensor reading prior to transmission and a post-processing function to the signal received by the fusion center (i.e., the sum of the individual transmit signals) to ensure a structural match between the function of interest and the wireless channel with its superposition property. As an immediate consequence, this enables the efficient estimation of functions of the form $f(s_1,\dots,s_N)=\psi(\sum_{i=1}^N\varphi_i(s_i))$, where $s_1,\dots,s_N$ denote the sensor readings and $\varphi_1,\dots,\varphi_N,\psi$ certain univariate functions. Even though \cite{Goldenbaum:Stanczak:13a} contains some interesting nonlinear function examples having such a representation, it lacks a comprehensive characterization of the corresponding function space. Reference \cite{Goldenbaum:Boche:Stanczak:13b} provides this characterization and points out that multivariate functions representable in the above manner are known as \emph{nomographic functions} \cite{Buck:79}. 

In contrast to the analog approach proposed in \cite{Goldenbaum:Stanczak:13a} (see \cite{Kortke:Goldenbaum:Stanczak:14} for a proof of concept), we present in this paper a simple digital scheme that extends the study of \cite{Goldenbaum:Boche:Stanczak:13b} to the reliable computation of nomographic functions in clustered Gaussian sensor networks.\footnote{By a clustered Gaussian sensor network we mean a clustered sensor network in which the intra-cluster communication takes place over Gaussian multiple-access channels.} The idea is as follows: each node in the network first quantizes its real-valued pre-processed sensor readings and then employs a nested lattice code from \cite{Nazer:Gastpar:11b} and \cite{Nazer:Dimakis:Gastpar:11} to protect the sum of messages against Gaussian channel noise. Decoding the sum and applying the corresponding post-processing function provides a reliable estimate of the sought function value.

It turns out that this combination of analog data pre- and post-processing with nested lattice codes allows for the computation of numerous nomographic functions at a computation rate that is not achievable with a separation-based method. The computation rate is thereby defined to be the number of function values that can be reliably computed per channel use. Furthermore, if some finite number of different nomographic functions is allowed to be computed over the channel one after another, then it is shown that every continuous function of the sensor readings can be treated. In addition to the improved rate performance, the proposed scheme has several other properties that are essential for wireless sensor network applications such as universality, lower decoding complexity, less coordination burden as well as the ability to deal with maximum decoding error probabilities.
%
%
%
%
%%%%%%%%%%%%%%%%%%%%%%%%%%%%%%%%%%%%%%%%%%%%%%%%%%%%%%%%%%%%%%%%%%%%%%%%%%%%%%%%%%%%%%%%%%%%%%%%%%%%%%%%%%%%%%%%%%%%%%%%%%%
\subsection{Related Work and Paper Organization} \label{sec:related}
Besides \cite{Nazer:Gastpar:07b} and our own prior work, the computation of special functions over a multiple-access channel (MAC) is considered for instance in \cite{Duman:Salehi:98}--\nocite{Mergen:Tong:06,Keller:Karamchandani:Fragouli:10}\cite{Tepedelenlioglu:Dasarathan:11}. To achieve performance gains, these works assume some \emph{structural match} between the function to be computed (e.g., an estimator or detector) and the operation the underlying MAC naturally performs. In this context, assuming arbitrary functions and arbitrary MACs, the authors of \cite{Karamchandani:Niesen:Diggavi:13} analyze the impact of structural mismatches on the computation performance. They show that for most pairs of functions and MACs a separation-based strategy is optimal. 

The majority of the above-referenced works implicitly deals with simple star-topology networks. The problem of computing functions over wireless networks with a more general topology is considered for instance in \cite{Zhan:Park:Gastpar:Sahai:13} and \cite{Wang:Jeon:Gastpar:13b}. Note that reference \cite{Goldenbaum:Boche:Stanczak:13b} considers a clustered network topology that is also used in this paper but under the assumption of \emph{noiseless} communication between nodes and fusion centers. Although this simplification provides insights for better understanding the mathematical subtleties of nonlinear computations over wireless networks (Section~\ref{sec:nomographic} contains a short summary), there is no coding scheme in \cite{Goldenbaum:Boche:Stanczak:13b}.

As already mentioned, based on the results in \cite{Erez:Zamir:04,Erez:Litsyn:Zamir:05}, Nazer and Gastpar propose in \cite{Nazer:Gastpar:11b} a lattice coding scheme that allows an efficient and reliable decoding of linear combinations of user messages in relay networks. Wilson et al. followed a similar approach in \cite{Wilson:Narayanan:Pfister:Sprintson:10} whereas the same setting is extended in \cite{Nokleby:Aazhang:12} to cooperating transmitters.

The paper is organized as follows. Section~\ref{sec:network} provides the network model and the problem statement. In Section~\ref{sec:nomographic}, the notion of nomographic functions is specified followed by some results to demonstrate that nomographic functions are well suited for distributed computation. Then, in Section~\ref{sec:reliable_computation}, we propose a corresponding computation scheme consisting of a novel data pre- and post-processing strategy along with a nested lattice computation code. Subsequently, Section~\ref{sec:achievable_rates} is devoted to determine the performance of the proposed scheme in terms of achievable computation rates as well as to comparisons with standard separation-based methods. Finally, Section~\ref{sec:conclusion} concludes the paper.  
%
%
%%%%%%%%%%%%%%%%%%%%%%%%%%%%%%%%%%%%%%%%%%%%%%%%%%%%%%%%%%%%%%%%%%%%%%%%%%%%%%%%%%%%%%%%%%%%%%%%%%%%%%%%%%%%%%%%%%%%%%%%%%%
\subsection{Notational Remarks} \label{sec:notation}
The natural, integer, real, and nonnegative real numbers are denoted as $\mathds{N}$, $\mathds{Z}$, $\mathds{R}$, and $\mathds{R}_+$ whereas $\mathds{E}\coloneqq [0,1]\subset\mathds{R}$ denotes the closed unit interval. For some $p\in\mathds{N}$, $\mathds{Z}_{p}=\{0,\dots,p-1\}$ denotes the integers modulo $p$ and $\bigoplus$ summation modulo $p$. The $n$-fold Cartesian product $\mathds{A}\times\cdots\times\mathds{A}$ of some set $\mathds{A}$ is written as $\mathds{A}^{n}$. Random variables are denoted by uppercase letters and their realizations by lowercase letters, respectively, whereas vectors are denoted by bold lowercase letters and matrices by bold uppercase letters. Let $\mathds{A}^{n}$ be some topological space, then $\C^0(\mathds{A}^{n})$ denotes the space of real-valued continuous functions with domain $\mathds{A}^{n}$. In contrast, $\F(\mathds{A}^{n})$ denotes the space of \emph{every} function $f:\mathds{A}^{n}\rightarrow\mathds{R}$. The volume of a closed subset $\mathds{D}$ of $\mathds{R}^{n}$ is described by $\Vol(\mathds{D})$, $\ma{I}_{n}$ denotes the $n\times n$ identity matrix, and $\log_2^+(x)\coloneqq\max\{\log_2(x),0\}$.
%
%
%
%
%%%%%%%%%%%%%%%%%%%%%%%%%%%%%%%%%%%%%%%%%%%%%%%%%%%%%%%%%%%%%%%%%%%%%%%%%%%%%%%%%%%%%%%%%%%%%%%%%%%%%%%%%%%%%%%%%%%%%%%%%%%
%%%%%%%%%%%%%%%%%%%%%%%%%%%%%%%%%%%%%%%%%%%%%%%%%%%%%%%%%%%%%%%%%%%%%%%%%%%%%%%%%%%%%%%%%%%%%%%%%%%%%%%%%%%%%%%%%%%%%%%%%%%
%	Network Model
%%%%%%%%%%%%%%%%%%%%%%%%%%%%%%%%%%%%%%%%%%%%%%%%%%%%%%%%%%%%%%%%%%%%%%%%%%%%%%%%%%%%%%%%%%%%%%%%%%%%%%%%%%%%%%%%%%%%%%%%%%%
%%%%%%%%%%%%%%%%%%%%%%%%%%%%%%%%%%%%%%%%%%%%%%%%%%%%%%%%%%%%%%%%%%%%%%%%%%%%%%%%%%%%%%%%%%%%%%%%%%%%%%%%%%%%%%%%%%%%%%%%%%%
%
\section{Network Model \& Problem Statement} \label{sec:network}
Consider a wireless sensor network consisting of $N\in\mathds{N}$ spatially distributed nodes that periodically monitor the environment resulting in a time series of sensor readings $\{s_i[t]\in\mathds{E}\}_{t\in\mathds{N}}$, $i=1,\dots,N$.\footnote{Assuming the sensor readings to be taken from the unit interval means no loss in generality as the results of this paper are valid for every compact interval of real numbers.} Assume that the network is organized into $L\in\mathds{N}$ clusters, where the set of nodes belonging to cluster $\ell$ is denoted by $C_{\ell}$, $\ell=1,\dots,L$. In particular, we focus on those clustered networks in which for each $\ell$ there exists at least one $\ell'\neq\ell$ such that $C_{\ell}\cap C_{\ell'}\neq\varnothing$. 

\begin{figure}[!t]
	\centering
	\begin{picture}(0,0)%
		\includegraphics{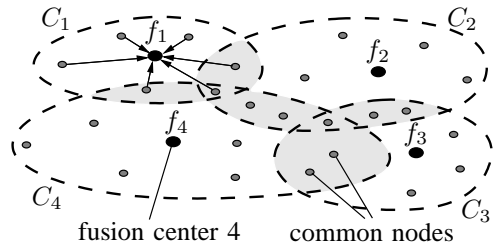}%
	\end{picture}%
	\setlength{\unitlength}{3947sp}%
	\begingroup\makeatletter\ifx\SetFigFont\undefined%
	\gdef\SetFigFont#1#2#3#4#5{%
	\reset@font\fontsize{#1}{#2pt}%
	\fontfamily{#3}\fontseries{#4}\fontshape{#5}%
	\selectfont}%
	\fi\endgroup%
	\begin{picture}(3052,1524)(3143,-1686)
		\put(3346,-318){\makebox(0,0)[lb]{\smash{{\SetFigFont{10}{12.0}{\rmdefault}{\mddefault}{\updefault}{\color[rgb]{0,0,0}$C_1$}%
		}}}}
		\put(3288,-1445){\makebox(0,0)[lb]{\smash{{\SetFigFont{10}{12.0}{\rmdefault}{\mddefault}{\updefault}{\color[rgb]{0,0,0}$C_4$}%
		}}}}
		\put(3578,-1667){\makebox(0,0)[lb]{\smash{{\SetFigFont{10}{12.0}{\rmdefault}{\mddefault}{\updefault}{\color[rgb]{0,0,0}fusion center 4}%
		}}}}
		\put(4908,-1671){\makebox(0,0)[lb]{\smash{{\SetFigFont{10}{12.0}{\rmdefault}{\mddefault}{\updefault}{\color[rgb]{0,0,0}common nodes}%
		}}}}
		\put(5983,-1527){\makebox(0,0)[lb]{\smash{{\SetFigFont{10}{12.0}{\rmdefault}{\mddefault}{\updefault}{\color[rgb]{0,0,0}$C_3$}%
		}}}}
		\put(5895,-318){\makebox(0,0)[lb]{\smash{{\SetFigFont{10}{12.0}{\rmdefault}{\mddefault}{\updefault}{\color[rgb]{0,0,0}$C_2$}%
		}}}}
		\put(3992,-399){\makebox(0,0)[lb]{\smash{{\SetFigFont{10}{12.0}{\rmdefault}{\mddefault}{\updefault}{\color[rgb]{0,0,0}$f_1$}%
		}}}}
		\put(5382,-520){\makebox(0,0)[lb]{\smash{{\SetFigFont{10}{12.0}{\rmdefault}{\mddefault}{\updefault}{\color[rgb]{0,0,0}$f_2$}%
		}}}}
		\put(5621,-1041){\makebox(0,0)[lb]{\smash{{\SetFigFont{10}{12.0}{\rmdefault}{\mddefault}{\updefault}{\color[rgb]{0,0,0}$f_3$}%
		}}}}
		\put(4117,-966){\makebox(0,0)[lb]{\smash{{\SetFigFont{10}{12.0}{\rmdefault}{\mddefault}{\updefault}{\color[rgb]{0,0,0}$f_4$}%
		}}}}
	\end{picture}%
	\caption{A clustered wireless sensor network consisting of $N=25$ nodes and $L=4$ clusters for computing some functions $f_1,\dots,f_4$ at FCs. Nodes belonging to one of the overlaps $C_{\ell}\cap C_{\ell'}$, $\ell\neq\ell'$, are called ``common nodes''.}%
	\label{fig:clustered_network}%
\end{figure}

Each cluster, consisting of $|C_{\ell}|$ nodes, has a designated fusion center (FC) that acts as the cluster head. Instead of reconstructing the sensor readings of all the assigned nodes, each FC aims at reliably and efficiently computing some given continuous function
\begin{equation}
	f_{\ell}:\mathds{E}^{|C_{\ell}|}\to\mathds{R}\;,\;\ve{s}_{\ell}[t]\mapsto f_{\ell}\bigl(\ve{s}_{\ell}[t]\bigr)\;,
	\label{eq:desired_function}
\end{equation}%
$\ell=1,\dots,L$, thereof, called \emph{desired function}. Here, the vector $\ve{s}_{\ell}[t]\in\mathds{E}^{|C_{\ell}|}$ contains all the sensor readings of cluster $\ell$. See Fig.~\ref{fig:clustered_network} for a qualitative example. 

Towards this end, each node, say node $i$, encodes its sensor readings into a length-$n$ sequence of transmit symbols, $x_i[1],\dots,x_i[n]$, subject to some average transmit power constraint $P>0$, that is, 
\begin{equation}%
	\sum_{m=1}^nx_i^2[m]\leq nP\;,\quad i=1,\dots,N\;.
	\label{eq:power_constraints}
\end{equation}%
To describe the intra-cluster communication between nodes and FCs, we use the standard discrete-time additive white Gaussian noise MAC (Gaussian MAC) so that the real-valued symbol received by FC $\ell$ at some channel use $m\in\{1,\dots,n\}$ is modeled as \cite{ElGamal:Kim:11}
\begin{equation}
	Y_{\ell}[m] = \sum_{i\in C_{\ell}}x_i[m] + Z_{\ell}[m]\;,\quad\ell=1,\dots,L\;.
	\label{eq:WMAC}
\end{equation}
Here and hereafter, $Z_{\ell}\sim\mathcal{N}(0,\sigma_Z^2)$ denotes for each $\ell$ independent and identically distributed additive white Gaussian noise (AWGN) with variance $\sigma_Z^2>0$. In all that follows, we call such a network a \emph{clustered Gaussian sensor network}.

Now, the problem to be solved in this paper is to efficiently compute at the FCs the desired functions $f_1,\dots,f_L$ with some pre-defined accuracy $\varepsilon>0$ by harnessing the superpositions in (\ref{eq:WMAC}). This is challenging due to the following reasons.\footnote{Note that the accuracy $\varepsilon$ as well as the desired functions are typically specified by the underlying sensor network application.}
\begin{description}
	\item[(i)] The common nodes can be heard by more than one FC, which results in interference between clusters. 
	\item[(ii)] The superposition of channel input symbols is corrupted by Gaussian noise.
\end{description}%
To account for these facts, we need to devise a novel computation scheme that combines an adapted data pre- and post-processing with a transmit strategy that fundamentally differs from those designed for typical message transfer. In particular, to address (ii), we employ a lattice code that is well suited for protecting sums of channel inputs whereas (i) is accounted for by exploiting the so-called \emph{universality property}, which is inherent to certain combinations of nomographic functions. The following section provides some mathematical background that helps to understand the results of this paper.  
%
%
%%%%%%%%%%%%%%%%%%%%%%%%%%%%%%%%%%%%%%%%%%%%%%%%%%%%%%%%%%%%%%%%%%%%%%%%%%%%%%%%%%%%%%%%%%%%%%%%%%%%%%%%%%%%%%%%%%%%%%%%%%%
%%%%%%%%%%%%%%%%%%%%%%%%%%%%%%%%%%%%%%%%%%%%%%%%%%%%%%%%%%%%%%%%%%%%%%%%%%%%%%%%%%%%%%%%%%%%%%%%%%%%%%%%%%%%%%%%%%%%%%%%%%%
%	Kolmogorov's Superpositions
%%%%%%%%%%%%%%%%%%%%%%%%%%%%%%%%%%%%%%%%%%%%%%%%%%%%%%%%%%%%%%%%%%%%%%%%%%%%%%%%%%%%%%%%%%%%%%%%%%%%%%%%%%%%%%%%%%%%%%%%%%%
%%%%%%%%%%%%%%%%%%%%%%%%%%%%%%%%%%%%%%%%%%%%%%%%%%%%%%%%%%%%%%%%%%%%%%%%%%%%%%%%%%%%%%%%%%%%%%%%%%%%%%%%%%%%%%%%%%%%%%%%%%%
%
\section{Nomographic Functions} \label{sec:nomographic}
Harnessing the superposition of different signals in (\ref{eq:WMAC}) for improving the network efficiency can be only beneficial if there is a \emph{structural match} between the channel operation and the desired function \cite{Nazer:Gastpar:07b,Karamchandani:Niesen:Diggavi:13}. A certain class of functions whose structure can properly be matched to channels that obey a superposition property are the so-called \emph{nomographic functions}, which are defined as follows \cite{Buck:79}, \cite{Goldenbaum:Boche:Stanczak:13b}.
\begin{definition}\label{def:nomographic}
	Let $N\geq 2$. Then, a function $f:\mathds{E}^N\to\mathds{R}$ 
for which there exist functions $\{\varphi_i\in\F(\mathds{E})\}_{i=1}^N$ and $\psi\in\F(\mathds{R})$ 
such that $f$ can be represented in the form
	\begin{equation}
		f(s_1,\dots,s_N)=\psi\left(\sum_{i=1}^N\varphi_i(s_i)\right)
		\label{eq:nomographic_function}
	\end{equation}%
	is called \emph{nomographic function}. The space of all nomographic functions with domain $\mathds{E}^N$ is denoted as $\Nomo(\mathds{E}^N)$.
\end{definition}%
Functions (\ref{eq:nomographic_function}) are called nomographic functions because they are the basis of nomographs, which are graphical aids for solving certain types of equations \cite{Epstein:58}. The following surprising result is due to Buck.
\begin{theorem}[Buck'79 \cite{Buck:79}]\label{thm:buck}
	Every function $f:\mathds{E}^N\to\mathds{R}$ is nomographic (i.e., $\Nomo(\mathds{E}^N)=\F(\mathds{E}^N)$).
\end{theorem}%
In what follows, we use $\Nomo^0(\mathds{E}^N)$ to denote the space of nomographic functions with the restriction that $\varphi_1,\dots,\varphi_N$ and $\psi$ have to be continuous. With this assumption in hand, Theorem~\ref{thm:buck} is no longer valid and we have the following result.
\begin{theorem}[Buck'82 \cite{Buck:82}]\label{thm:buck2}
	The space of nomographic functions with continuous pre- and post-processing functions is nowhere dense in the space of continuous functions, that is, $\Nomo^0(\mathds{E}^N)$ nowhere
dense in $\C^0(\mathds{E}^N)$.
\end{theorem}%
Theorem~\ref{thm:buck2} may appear discouraging but the following groundbreaking theorem provides a kind of remedy; it states that \emph{every} continuous multivariate function is representable as a simple sum of nomographic functions taken from $\Nomo^0(\mathds{E}^N)$.
\begin{theorem}[Kolmogorov'57 \cite{Kolmogorov:57}]\label{thm:kolmogorov}
	Every $f\in\C^0(\mathds{E}^N)$ can be represented as the superposition of at most $2N+1$ nomographic functions. That is, in the form
	\begin{equation}
		f(s_1,\dots,s_N)=\sum_{j=1}^{2N+1}g_j(s_1,\dots,s_N)\;,
		\label{eq:kolmogorov}
	\end{equation}%
	with $g_j(s_1,\dots,s_N)=\psi_j\bigl(\sum_{i=1}^N\varphi_{ij}(s_i)\bigr)\in\Nomo^0(\mathds{E}^N)$, in which only the $\psi_j\in\C^0(\mathds{R})$ depend on $f$ but the $N(2N+1)$ functions $\varphi_{ij}\in\C^0(\mathds{E})$ do not.
\end{theorem}%
\begin{remark}\label{rem:kolmogorov_superpositions}
	The theorem states that every continuous multivariate function can be represented as a superposition of only one variable functions. A fact that was claimed to be impossible by David Hilbert in the 13\textsuperscript{th} of his famous 23 problems stated in 1900 \cite{Hilbert:1902}. Representations (\ref{eq:kolmogorov}) are called \emph{Kolmogorov's superpositions}.
\end{remark}
According to Theorem~\ref{thm:kolmogorov}, $2N+1$ nomographic functions are sufficient to write every $f\in\C^0(\mathds{E}^N)$ in the form of (\ref{eq:kolmogorov}). Theorem~\ref{thm:sternfeld} strengthens this result. 
\begin{theorem}[Sternfeld'85 \cite{Sternfeld:85}]\label{thm:sternfeld}
	To represent every $f\in\C^0(\mathds{E}^N)$ as a Kolmogorov's superposition with elements from $\Nomo^0(\mathds{E}^N)$, there are at least $2N+1$ nomographic functions necessary (i.e., $2N+1$ cannot be reduced).
\end{theorem}%
\begin{remark}\label{rem:geometry}
	A geometric interpretation of Theorem~\ref{thm:kolmogorov}, which will be useful for the discussions in Section~\ref{sec:discussion}, is the following. Using in (\ref{eq:kolmogorov}) $2N+1$ inner sums results in a continuous and bijective correspondence $(s_1,\dots,s_N)\mapsto(\sum_i\varphi_{i1}(s_i),\dots,\sum_i\varphi_{i,2N+1}(s_i))\in\Gamma$, with $\Gamma$ being a compact subset of $\mathds{R}^{2N+1}$. In other words, $(\sum_i\varphi_{i1}(s_i),\dots,\sum_i\varphi_{i,2N+1}(s_i))$ describes a homeomorphism that embeds $\mathds{E}^N$ in $\mathds{R}^{2N+1}$.
\end{remark}%

Let $\varphi_{ij}$, $i=1,\dots,N$; $j=1,\dots,2N+1$, be $N(2N+1)$ continuous functions such that according to Theorems \ref{thm:kolmogorov} and \ref{thm:sternfeld} every continuous $f:\mathds{E}^N\to\mathds{R}$ can be written as
\begin{equation*}
	f(s_1,\dots,s_N)=\sum_{j=1}^{2N+1}\psi_{j}\left(\sum_{i=1}^N\varphi_{ij}(s_i)\right)
\end{equation*}%
through a proper choice of the continuous functions $\psi_{1},\dots,\psi_{2N+1}$. Suppose that each node in the network is uniquely assigned one of the function sets $\{\varphi_{ij}\in\C^0(\mathds{E})\}_{j=1}^{2N+1}$, $i=1,\dots,N$. In what follows, we call them the \emph{pre-processing functions} and point out that by Theorem \ref{thm:kolmogorov}, they do not depend on the choice of $f$. Then, the desired function (\ref{eq:desired_function}) to be computed at FC $\ell$, $\ell=1,\dots,L$, can be written as \cite{Goldenbaum:Boche:Stanczak:13b}
\begin{equation}
	f_{\ell}\bigl(\ve{s}_{\ell}[t]\bigr)=\sum_{j=1}^{2N+1}\psi_{\ell j}\Biggl(\sum_{i\in C_{\ell}}\varphi_{ij}\bigl(s_i[t]\bigr) + \gamma_{\ell j}\Biggr)
	\label{eq:kolmogorov_cluster_alt}
\end{equation}%
with the constants
\begin{equation}
	\gamma_{\ell j}\coloneqq\sum_{i\notin C_{\ell}}\varphi_{ij}(0)\;,\;j=1,\dots,2N+1\;,
	\label{eq:constants}
\end{equation}%
and appropriately chosen functions $\{\psi_{\ell j}\in\C^0(\mathds{R})\}_{j=1}^{2N+1}$, referred to as the \emph{post-processing functions}. Assume that FC $\ell$ knows its set of post-processing functions, $\{\psi_{\ell j}\}_{j=1}^{2N+1}$, as well as its set of constants $\{\gamma_{\ell j}\}_{j=1}^{2N+1}$. Then, the computation approach considered in this paper can be briefly outlined as follows:
\begin{itemize}
	\item[(i)] With sensor nodes transmitting concurrently in the same frequency band, the FCs reliably reconstruct for every $t\in\mathds{N}$ the sequences $\{\sum_{i\in C_{\ell}}\varphi_{ij}(s_i[t])\}_{j=1}^{2N+1}$, $\ell=1,\dots,L$, of superimposed pre-processed sensor readings. 
	\item[(ii)] The FCs add the constants (\ref{eq:constants}), apply their post-processing functions $\{\psi_{\ell j}\}_{j=1}^{2N+1}$ and sum up all intermediate results to yield the desired function values (\ref{eq:kolmogorov_cluster_alt}).
\end{itemize}%
\begin{remark}\label{rem:all_functions}
	Due to the fact that the pre-processing functions are independent of the functions to be computed at the FCs (see Theorem~\ref{thm:kolmogorov}), they do not need to be updated if the desired functions change during network operation. A property to which we refer as \emph{universality} \cite{Goldenbaum:Boche:Stanczak:13b}. The constants (\ref{eq:constants}) are for instance responsible for preserving the universality in (\ref{eq:kolmogorov_cluster_alt}) as otherwise, the pre-processing functions would depend on the cluster index $\ell$. See Section~\ref{sec:kolmogorov_multiple} for a detailed discussion.
\end{remark}%
%
%
%
%%%%%%%%%%%%%%%%%%%%%%%%%%%%%%%%%%%%%%%%%%%%%%%%%%%%%%%%%%%%%%%%%%%%%%%%%%%%%%%%%%%%%%%%%%%%%%%%%%%%%%%%%%%%%%%%%%%%%%%%%%%
%%%%%%%%%%%%%%%%%%%%%%%%%%%%%%%%%%%%%%%%%%%%%%%%%%%%%%%%%%%%%%%%%%%%%%%%%%%%%%%%%%%%%%%%%%%%%%%%%%%%%%%%%%%%%%%%%%%%%%%%%%%
%	Reliable Computation Over Wireless MACs
%%%%%%%%%%%%%%%%%%%%%%%%%%%%%%%%%%%%%%%%%%%%%%%%%%%%%%%%%%%%%%%%%%%%%%%%%%%%%%%%%%%%%%%%%%%%%%%%%%%%%%%%%%%%%%%%%%%%%%%%%%%
%%%%%%%%%%%%%%%%%%%%%%%%%%%%%%%%%%%%%%%%%%%%%%%%%%%%%%%%%%%%%%%%%%%%%%%%%%%%%%%%%%%%%%%%%%%%%%%%%%%%%%%%%%%%%%%%%%%%%%%%%%%
%
\section{Reliable Computation of Nomographic Functions Over the Gaussian MAC} \label{sec:reliable_computation}
\begin{figure*}[!t]
	\centering
	\begin{picture}(0,0)%
		\includegraphics{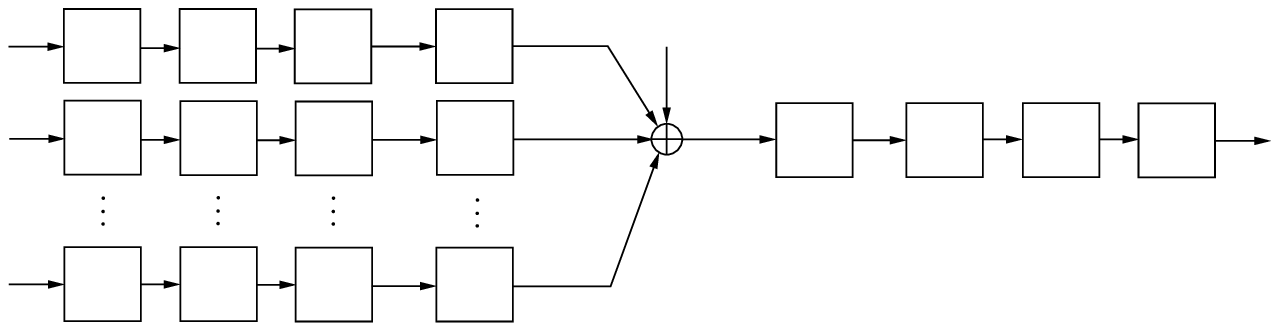}%
	\end{picture}%
	\setlength{\unitlength}{3947sp}%
	\begingroup\makeatletter\ifx\SetFigFont\undefined%
	\gdef\SetFigFont#1#2#3#4#5{%
	\reset@font\fontsize{#1}{#2pt}%
	\fontfamily{#3}\fontseries{#4}\fontshape{#5}%
	\selectfont}%
	\fi\endgroup%
	\begin{picture}(6213,1542)(855,-5866)
		\put(1372,-4552){\makebox(0,0)[lb]{\smash{{\SetFigFont{9}{10.8}{\rmdefault}{\mddefault}{\updefault}{\color[rgb]{0,0,0}$\varphi_1$}%
		}}}}
		\put(1370,-4993){\makebox(0,0)[lb]{\smash{{\SetFigFont{9}{10.8}{\rmdefault}{\mddefault}{\updefault}{\color[rgb]{0,0,0}$\varphi_2$}%
		}}}}
		\put(1356,-5694){\makebox(0,0)[lb]{\smash{{\SetFigFont{9}{10.8}{\rmdefault}{\mddefault}{\updefault}{\color[rgb]{0,0,0}$\varphi_N$}%
		}}}}
		\put(885,-4477){\makebox(0,0)[lb]{\smash{{\SetFigFont{9}{10.8}{\rmdefault}{\mddefault}{\updefault}{\color[rgb]{0,0,0}$s_1[t]$}%
		}}}}
		\put(887,-4917){\makebox(0,0)[lb]{\smash{{\SetFigFont{9}{10.8}{\rmdefault}{\mddefault}{\updefault}{\color[rgb]{0,0,0}$s_2[t]$}%
		}}}}
		\put(870,-5617){\makebox(0,0)[lb]{\smash{{\SetFigFont{9}{10.8}{\rmdefault}{\mddefault}{\updefault}{\color[rgb]{0,0,0}$s_N[t]$}%
		}}}}
		\put(1937,-4574){\makebox(0,0)[lb]{\smash{{\SetFigFont{9}{10.8}{\rmdefault}{\mddefault}{\updefault}{\color[rgb]{0,0,0}$Q$}%
		}}}}
		\put(1945,-5017){\makebox(0,0)[lb]{\smash{{\SetFigFont{9}{10.8}{\rmdefault}{\mddefault}{\updefault}{\color[rgb]{0,0,0}$Q$}%
		}}}}
		\put(1949,-5718){\makebox(0,0)[lb]{\smash{{\SetFigFont{9}{10.8}{\rmdefault}{\mddefault}{\updefault}{\color[rgb]{0,0,0}$Q$}%
		}}}}
		\put(6548,-5013){\makebox(0,0)[lb]{\smash{{\SetFigFont{9}{10.8}{\rmdefault}{\mddefault}{\updefault}{\color[rgb]{0,0,0}$\psi$}%
		}}}}
		\put(6809,-4934){\makebox(0,0)[lb]{\smash{{\SetFigFont{9}{10.8}{\rmdefault}{\mddefault}{\updefault}{\color[rgb]{0,0,0}$\hat{f}$}%
		}}}}
		\put(5918,-5016){\makebox(0,0)[lb]{\smash{{\SetFigFont{9}{10.8}{\rmdefault}{\mddefault}{\updefault}{\color[rgb]{0,0,0}$Q^{-1}$}%
		}}}}
		\put(3430,-5650){\makebox(0,0)[lb]{\smash{{\SetFigFont{9}{10.8}{\rmdefault}{\mddefault}{\updefault}{\color[rgb]{0,0,0}$\ve{x}_N$}%
		}}}}
		\put(3437,-4944){\makebox(0,0)[lb]{\smash{{\SetFigFont{9}{10.8}{\rmdefault}{\mddefault}{\updefault}{\color[rgb]{0,0,0}$\ve{x}_2$}%
		}}}}
		\put(3430,-4497){\makebox(0,0)[lb]{\smash{{\SetFigFont{9}{10.8}{\rmdefault}{\mddefault}{\updefault}{\color[rgb]{0,0,0}$\ve{x}_1$}%
		}}}}
		\put(4263,-4931){\makebox(0,0)[lb]{\smash{{\SetFigFont{9}{10.8}{\rmdefault}{\mddefault}{\updefault}{\color[rgb]{0,0,0}$\ve{y}$}%
		}}}}
		\put(4780,-5029){\makebox(0,0)[lb]{\smash{{\SetFigFont{9}{10.8}{\rmdefault}{\mddefault}{\updefault}{\color[rgb]{0,0,0}$\mathcal{D}_2$}%
		}}}}
		\put(5072,-4935){\makebox(0,0)[lb]{\smash{{\SetFigFont{9}{10.8}{\rmdefault}{\mddefault}{\updefault}{\color[rgb]{0,0,0}$\hat{\ve{g}}$}%
		}}}}
		\put(3173,-5724){\makebox(0,0)[lb]{\smash{{\SetFigFont{9}{10.8}{\rmdefault}{\mddefault}{\updefault}{\color[rgb]{0,0,0}$\mathcal{E}_2$}%
		}}}}
		\put(3171,-5018){\makebox(0,0)[lb]{\smash{{\SetFigFont{9}{10.8}{\rmdefault}{\mddefault}{\updefault}{\color[rgb]{0,0,0}$\mathcal{E}_2$}%
		}}}}
		\put(3167,-4578){\makebox(0,0)[lb]{\smash{{\SetFigFont{9}{10.8}{\rmdefault}{\mddefault}{\updefault}{\color[rgb]{0,0,0}$\mathcal{E}_2$}%
		}}}}
		\put(2756,-4498){\makebox(0,0)[lb]{\smash{{\SetFigFont{9}{10.8}{\rmdefault}{\mddefault}{\updefault}{\color[rgb]{0,0,0}$\ve{w}_{1}$}%
		}}}}
		\put(2758,-4944){\makebox(0,0)[lb]{\smash{{\SetFigFont{9}{10.8}{\rmdefault}{\mddefault}{\updefault}{\color[rgb]{0,0,0}$\ve{w}_2$}%
		}}}}
		\put(2754,-5642){\makebox(0,0)[lb]{\smash{{\SetFigFont{9}{10.8}{\rmdefault}{\mddefault}{\updefault}{\color[rgb]{0,0,0}$\ve{w}_N$}%
		}}}}
		\put(4105,-4480){\makebox(0,0)[lb]{\smash{{\SetFigFont{9}{10.8}{\rmdefault}{\mddefault}{\updefault}{\color[rgb]{0,0,0}$\ve{z}$}%
		}}}}
		\put(2496,-4578){\makebox(0,0)[lb]{\smash{{\SetFigFont{9}{10.8}{\rmdefault}{\mddefault}{\updefault}{\color[rgb]{0,0,0}$\mathcal{E}_1$}%
		}}}}
		\put(2496,-5024){\makebox(0,0)[lb]{\smash{{\SetFigFont{9}{10.8}{\rmdefault}{\mddefault}{\updefault}{\color[rgb]{0,0,0}$\mathcal{E}_1$}%
		}}}}
		\put(2495,-5720){\makebox(0,0)[lb]{\smash{{\SetFigFont{9}{10.8}{\rmdefault}{\mddefault}{\updefault}{\color[rgb]{0,0,0}$\mathcal{E}_1$}%
		}}}}
		\put(5412,-5025){\makebox(0,0)[lb]{\smash{{\SetFigFont{9}{10.8}{\rmdefault}{\mddefault}{\updefault}{\color[rgb]{0,0,0}$\mathcal{D}_1$}%
		}}}}
	\end{picture}%
	\caption{Block diagram of the entire transmission chain when computing a nomographic function over a Gaussian multiple-access channel with $N$ nodes (i.e., within a single cluster).}
	\label{fig:nomographic}%
\end{figure*}%
What nomographic functions makes so interesting for our considerations is the mentioned structural match between the inner sums and the channel operation (i.e., superposition), which suggests that harnessing these signal interactions may has the potential for improving the efficiency in wireless sensor networks also for computing nonlinear functions. Realizing this in a reliable manner requires the application of coding techniques as we have to deal with additive noise. This leads us to the formal notion of a computation code \cite{Nazer:Gastpar:07b}.
\begin{definition}\label{def:computation_code}
	Let $T,n\in\mathds{N}$ be chosen arbitrarily. An $(f,T,n)$ \emph{computation code} for a Gaussian MAC consists of:
	\begin{itemize}
		\item A desired function $f\in\F(\mathds{E}^N)$.
		\item $N$ encoding functions that map $T$ sensor readings $s_i[1],\dots,s_i[T]$ to $n$ channel input symbols each.
		\item A decoding function that assigns $T$ estimates $\hat{f}(\ve{s}[1]),\dots,\hat{f}(\ve{s}[T])$ of desired function-values to each length-$n$ sequence of channel output symbols.
	\end{itemize}%
\end{definition}%
The performance of a computation code is typically determined in terms of an achievable computation rate, which specifies how many function values can be computed per channel use within a predefined accuracy.
\begin{definition}\label{def:computation_rate}
	Let $f\in\F(\mathds{E}^N)$ be some fixed desired function, $\hat{f}$ a corresponding estimate at the FC, and $\varepsilon>0$ an arbitrary but fixed computation accuracy. Then, $R^{\text{Comp}}(f,\varepsilon)\in\mathds{R}_+$ (in function-values per channel use) is said to be an \emph{achievable computation rate} for $f$ and $\varepsilon$ if for every rate $R'\coloneqq\frac{T}{n}<R^{\text{Comp}}(f,\varepsilon)$ and every $\delta>0$ there exists an $(f,nR',n)$ computation code such that the error probability fulfills
\begin{equation}
	\Prob\left(\bigcup_{t=1}^{T}\left\{\sup_{\ve{s}[t]\in\mathds{E}^N}\Bigl|\hat{f}\bigl(\ve{s}[t]\bigr)-f\bigl(\ve{s}[t]\bigr)\Bigr|>\varepsilon\right\}\right)<\delta\;,
	\label{eq:computation_rate_error}
\end{equation}%
for $n$ sufficiently large.% as long as the coding rate (\ref{eq:message_rate}) fulfills at each node $R<R^{\text{Comp}}$.
\end{definition}%

For ease of exposition, in the following two subsections we consider the single cluster case (i.e., $L=1$), whereas the extension to the general multiple cluster case is considered in Section~\ref{sec:multiple_cluster} below. In particular, we propose a computation code for some fixed continuous nomographic function that is capable of achieving (\ref{eq:computation_rate_error}) at computation rates that are, to some extent, not achievable with separation-based strategies. Since the encoders and the decoder have to respect the particular structure of nomographic functions, we decompose them into multiple components each. 
%
%
%
%%%%%%%%%%%%%%%%%%%%%%%%%%%%%%%%%%%%%%%%%%%%%%%%%%%%%%%%%%%%%%%%%%%%%%%%%%%%%%%%%%%%%%%%%%%%%%%%%%%%%%%%%%%%%%%%%%%%%%%%%%%
\subsection{Data Pre- and Post-Processing} \label{sec:processing}
Let $f\in\Nomo^0(\mathds{E}^N)$ so that it can be expressed as 
\begin{equation}
	f\bigl(s_1[t],\dots,s_N[t]\bigr)=\psi\left(\sum_{i=1}^N\varphi_i\bigl(s_i[t]\bigr)\right)
	\label{eq:nomographic2}
\end{equation}%
through a proper choice of continuous pre- and post-processing functions. One of the basic facts in multiuser information theory states that the Gaussian MAC in (\ref{eq:WMAC}) is a \emph{finite capacity} channel if transmit powers and bandwidths are finite \cite{ElGamal:Kim:11}. As a consequence, communicating the real values $\varphi_i(s_i[t])$ over such a channel with infinite precision is not possible so that we have to first quantize the pre-processed sensor readings into binary representations. Since $\mathds{E}$ is a compact interval, the range of each pre-processing function is a compact interval as well and we denote these sets by $\Pi_{i}\subset\mathds{R}$ (i.e., $\forall s\in\mathds{E}:\varphi_{i}(s)\in\Pi_{i}$). Thus, it follows that the union $\Pi\coloneqq\bigcup_{i=1}^N\Pi_{i}$ is a compact interval and we denote by $\pi_{\text{max}}\coloneqq\max_{\xi\in\Pi}|\xi|$ the unique maximal element in absolute value. 
\begin{remark}\label{rem:nonnegativity}
	To keep the notation simple, we assume in the following that the elements of $\Pi$ are nonnegative. This is without loss of generality as $\Pi$ can be shifted to the nonnegative reals by adding $\pi_{\text{max}}$ to every $\xi\in\Pi$.
\end{remark}%

Each node in the network employs the same \emph{quantizer}
\begin{equation}
	Q:\Pi\to\{0,1\}^b
	\label{eq:quantizer}
\end{equation}%
in order to form for every $t\in\mathds{N}$ the length-$b$ binary representation
\begin{equation}
	\ve{w}_{i}[t]\coloneqq Q\bigl(\varphi_{i}(s_i[t])\bigr)\;,\;i=1,\dots,N\;,
	\label{eq:messages}
\end{equation}%
where $b$ is some positive integer to be specified below. To better understand how quantizer $Q$ works, recall first from \cite[Thm. 5.2]{Kulisch:08} that every $\xi\in\Pi$ has a unique dyadic expansion
\begin{equation*}
	\xi=\sum_{r=-v}^{\infty}\frac{w_r}{2^r}=\lim_{\eta\to\infty}\sum_{r=-v}^{\eta}\frac{w_r}{2^r}\;,
	\label{eq:real_representation}
\end{equation*}%
with $w_r\in\{0,1\}$ and $w_r\neq 1$ for infinitely many $r$, unless $w_r=1$ for all $r$. Observe that $v$ depends on the largest integer part of $\xi$. With this in mind, consider for each $i$, $i=1,\dots,N$, the instantaneous approximation
\begin{equation}
	\varphi_{i}\bigl(s_i[t]\bigr)\approx\tilde{\varphi}_{i}\bigl(s_i[t]\bigr)=\sum_{r=-v}^{\eta} w_{ri}[t]2^{-r}
	\label{eq:approx_phi}
\end{equation}%
by terminating the dyadic expansion. Then, setting $b\coloneqq \eta+v+1$ with $v\coloneqq\lfloor\log_2(\pi_{\text{max}})\rfloor$ fixed, quantizer $Q$ simply forms the length-$b$ representations in (\ref{eq:messages}) by extracting the binary digits from (\ref{eq:approx_phi}).  

Each quantizer is followed by the same \emph{source encoder}
\begin{equation}
	\mathcal{E}_1:\{0,1\}^{bT}\to\mathds{Z}_p^k\;,
	\label{eq:source_encoder}
\end{equation}%
which combines $T\in\mathds{N}$ of the binary representations to a length-$k$ message over $\mathds{Z}_p$:
\begin{equation}
	\ve{w}_i=\mathcal{E}_1\bigl(\ve{w}_i[1],\dots,\ve{w}_i[T]\bigr)\;,\;i=1,\dots,N\;.
	\label{eq:overall_messages}
\end{equation}%
Here and hereafter, $k$ is a natural number and $p$ is assumed to be prime.\footnote{We construct the encoder (\ref{eq:source_encoder}) explicitly in the proof of Theorem~\ref{thm:sum_protection}.} See Fig.~\ref{fig:nomographic} for a block diagram.

Now, in order to compute the desired function in (\ref{eq:nomographic2}) over the Gaussian MAC (\ref{eq:WMAC}), the FC first needs for every fixed $t$, $t=1,\dots,T$, a reliable estimate of the inner sum
\begin{equation}
	\tilde{g}\bigl(s_1[t],\dots,s_N[t]\bigr)\coloneqq\sum_{i=1}^N\tilde{\varphi}_{i}\bigl(s_i[t]\bigr)\;.
	\label{eq:quantized_sum_pre}
\end{equation}%
This can be achieved by reliably computing the \emph{modulo $p$ sum} of the messages (\ref{eq:overall_messages}),
\begin{equation}
	\ve{g}\coloneqq\bigoplus_{i=1}^N\ve{w}_{i}\;,
	\label{eq:mod_b_messages}
\end{equation}%
as long as $p$ is sufficiently large.
\begin{remark}\label{rem:wraparound}
	Requiring $p$ to be sufficiently large is necessary in order to avoid a wraparound in (\ref{eq:mod_b_messages}).
\end{remark}%

Once the FC knows (\ref{eq:mod_b_messages}), a \emph{source decoder}
\begin{equation*}
	\mathcal{D}_1:\mathds{Z}_p^k\to\{0,\dots,N\}^{bT}
\end{equation*}%
first decomposes $\ve{g}$ into modulo $p$ sums of the binary representations (\ref{eq:messages}):
\begin{equation}
	\mathcal{D}_1(\ve{g})=\bigl(\ve{g}[1],\dots,\ve{g}[T]\bigr)\coloneqq\left(\bigoplus_{i=1}^N\ve{w}_i[1],\dots,\bigoplus_{i=1}^N\ve{w}_i[T]\right)\;.
	\label{eq:decoder_1}
\end{equation}%
Afterwards, the inverse quantizer evaluates the right hand side of (\ref{eq:approx_phi}) at the digits $\ve{g}[t]$, $t=1,\dots,T$, followed by the post-processing function $\psi$ (see Fig.~\ref{fig:nomographic}). This data post-processing provides the FC with an approximation of (\ref{eq:nomographic2}) given by
\begin{equation}
	\tilde{f}\bigl(s_1[t],\dots,s_N[t]\bigr)
	\coloneqq\psi\left(\sum_{i=1}^N\tilde{\varphi}_{i}\bigl(s_i[t]\bigr)\right)\;.
	\label{eq:f_approx}
\end{equation}%
Note that the corresponding accuracy crucially depends on the explicit choice of the quantization parameter $b$. As this will also have a significant impact on the achievable computation rate, we provide in Lemma~\ref{lem:approximation_error} at the beginning of Section~\ref{sec:achievable_rates} the relationship between $b$ and some given accuracy $\varepsilon>0$.
%
%
%
%%%%%%%%%%%%%%%%%%%%%%%%%%%%%%%%%%%%%%%%%%%%%%%%%%%%%%%%%%%%%%%%%%%%%%%%%%%%%%%%%%%%%%%%%%%%%%%%%%%%%%%%%%%%%%%%%%%%%%%%%%%
\subsection{Nested Lattice Coding} \label{sec:lattice_codes}
%
%
%%%%%%%%%%%%%%%%%%%%%%%%%%%%%%%%%%%%%%%%%%%%%%%%%%%%%%%%%%%%%%%%%%%%%%%%%%%%%%%%%%%%%%%%%%%%%%%%%%%%%%%%%%%%%%%%%%%%%%%%%%%
\subsubsection{Basic Facts and Definitions} \label{sec:lattices_facts}
Reading through Section~\ref{sec:processing} reveals that the crucial step in achieving reliable computations over the channel is the protection of (\ref{eq:mod_b_messages}) against Gaussian noise. In order to ensure this, we employ sequences of lattice codes proposed in \cite{Nazer:Gastpar:11b} as they possess favorable structural properties. Towards this end, we first briefly recap some necessary notions on nested lattices from \cite{Erez:Zamir:04,Erez:Litsyn:Zamir:05,ForneyJr:03}.

An $n$-dimensional \emph{lattice} $\Lambda$, $n\in\mathds{N}$, is a discrete additive subgroup of $\mathds{R}^n$ that is closed under addition. For every $\Lambda$ there exists a full-rank generator matrix $\ma{G}\in\mathds{R}^{n\times n}$ so that
	\begin{equation*}
		\Lambda=\{\ve{\lambda}=\ma{G}\ve{\mu}\,|\,\ve{\mu}\in\mathds{Z}^n\}\eqqcolon\ma{G}\mathds{Z}^n\;.
	\end{equation*}%
A \emph{quantizer} associated with $\Lambda$ is a map $Q_{\Lambda}:\mathds{R}^n\to\Lambda$ that assigns every $\ve{\mu}\in\mathds{R}^n$ to the nearest lattice point in Euclidean distance, that is,
	\begin{equation*}
		Q_{\Lambda}(\ve{\mu})=\argmin_{\ve{\lambda}\in\Lambda}\|\ve{\mu}-\ve{\lambda}\|_2\;.
	\end{equation*}%
The \emph{fundamental Voronoi region} of $\Lambda$, denoted as $\mathcal{V}$, is the
set of all points that quantize to the zero vector:
		\begin{equation*}
			\mathcal{V}\coloneqq\{\ve{\mu}\in\mathds{R}^n\,|\,Q_{\Lambda}(\ve{\mu})=\ve{0}\}\;.
		\end{equation*}%
The \emph{modulo operation} with respect to $\Lambda$ provides for every $\ve{\mu}\in\mathds{R}^n$ the quantization error, which is always in $\mathcal{V}$:
		\begin{equation*}
			[\ve{\mu}]\bmod{\Lambda}\coloneqq\ve{\mu}-Q_{\Lambda}(\ve{\mu})\;.
		\end{equation*}%
\begin{definition}\label{def:second_moment}
	The \emph{second moment} (per dimension) of a given lattice $\Lambda\subset\mathds{R}^n$ is defined as
		\begin{equation}
			\sigma^2(\Lambda)\coloneqq\frac{1}{n}\frac{\int_{\mathcal{V}}\|\ve{x}\|_2^2\,\textrm{d}\ve{x}}{\Vol(\mathcal{V})}\;,
			\label{eq:2nd_moment}
		\end{equation}%
	with $\Vol(\mathcal{V})=\int_{\mathcal{V}}\textrm{d}\ve{x}$ the volume of the fundamental Voronoi region. The \emph{normalized second moment} is defined as
		\begin{equation}%
			G(\Lambda)\coloneqq\frac{\sigma^2(\Lambda)}{\Vol(\mathcal{V})^{2/n}}\;.
			\label{eq:normalized_2nd_moment}
		\end{equation}%	
\end{definition}%
\begin{definition}\label{def:goodness}
	Let $\{\Lambda^{(n)}\}$ be a sequence of lattices indexed by their dimension and $\ve{z}\sim\mathcal{N}(\ve{0},\sigma_Z^2\ma{I}_n)$. Then, $\{\Lambda^{(n)}\}$ is said to be \emph{good for AWGN channel coding} if $\Prob(\ve{z}\notin\mathcal{V}^{(n)})\to 0$ exponentially fast with growing $n$ whenever $\Vol(\mathcal{V}^{(n)})^{2/n}>2\pi\!\e\sigma_Z^2$ applies. On the other hand, $\{\Lambda^{(n)}\}$ is said to be \emph{good for shaping} if $\lim_{n\to\infty}\log_2(2\pi\!\e G(\Lambda^{(n)}))=0$.
\end{definition}%

\begin{figure}[!t]
	\centering
	\begin{picture}(0,0)%
	\includegraphics{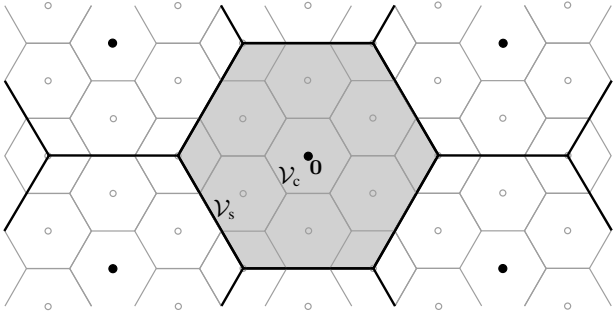}%
	\end{picture}%
	\setlength{\unitlength}{3947sp}%
	\begingroup\makeatletter\ifx\SetFigFont\undefined%
	\gdef\SetFigFont#1#2#3#4#5{%
	\reset@font\fontsize{#1}{#2pt}%
	\fontfamily{#3}\fontseries{#4}\fontshape{#5}%
	\selectfont}%
	\fi\endgroup%
	\begin{picture}(3859,1949)(1445,-2261)
		\put(3186,-1459){\makebox(0,0)[lb]{\smash{{\SetFigFont{9}{10.8}{\rmdefault}{\mddefault}{\updefault}{\color[rgb]{0,0,0}$\mathcal{V}_{\text{c}}$}%
		}}}}
		\put(2781,-1667){\makebox(0,0)[lb]{\smash{{\SetFigFont{9}{10.8}{\rmdefault}{\mddefault}{\updefault}{\color[rgb]{0,0,0}$\mathcal{V}_{\text{s}}$}%
		}}}}
		\put(3384,-1400){\makebox(0,0)[lb]{\smash{{\SetFigFont{8}{9.6}{\rmdefault}{\mddefault}{\updefault}{\color[rgb]{0,0,0}$\ve{0}$}%
		}}}}
	\end{picture}%
	\caption{Part of a nested hexagonal lattice $\Lambda_{\text{s}}\subset\Lambda_{\text{c}}$ in Euclidean space $\mathds{R}^2$ with $\mathcal{V}_{\text{s}}$ the fundamental Voronoi region of the shaping lattice $\Lambda_{\text{s}}$ (black dots) and $\mathcal{V}_{\text{c}}$ the fundamental Voronoi region of the coding lattice $\Lambda_{\text{c}}$ (white dots).}
	\label{fig:nested_lattice}
\end{figure}

A lattice $\Lambda_{\text{s}}$ is \emph{nested} in some lattice $\Lambda_{\text{c}}$ if $\Lambda_{\text{s}}\subset\Lambda_{\text{c}}$. Here and hereafter, $\Lambda_{\text{s}}$ with fundamental Voronoi region $\mathcal{V}_{\text{s}}$ is called \emph{shaping lattice} whereas $\Lambda_{\text{c}}$ with fundamental Voronoi region $\mathcal{V}_{\text{c}}$ is called \emph{coding lattice}. Fig.~\ref{fig:nested_lattice} depicts an example of a nested hexagonal lattice pair in which $\Lambda_{\text{c}}=\ma{G}\mathds{Z}^2$ and $\Lambda_{\text{s}}=3\ma{G}\mathds{Z}^2$, with generator matrix $\ma{G}=\left(\begin{smallmatrix}\sqrt{3}/2 &0\\ 1/2 &1 \end{smallmatrix}\right)$.
\begin{lemma}\label{lem:modulus}
	For all $\ve{\mu},\ve{\nu}\in\mathds{R}^n$ and every pair of nested lattices $\Lambda\subset\Lambda'$, the modulo operation satisfies:
		\begin{align}
			[\ve{\mu}+\ve{\nu}]\bmod{\Lambda}&=\bigl[[\ve{\mu}]\bmod{\Lambda}+\ve{\nu}\bigr]\bmod{\Lambda}\label{eq:dist}\\
			[Q_{\Lambda'}(\ve{\mu})]\bmod{\Lambda}&=[Q_{\Lambda'}([\ve{\mu}]\bmod{\Lambda})]\bmod{\Lambda}\label{eq:comm}\;.
		\end{align}%
\end{lemma}%
\begin{IEEEproof}
	The proof is straightforward.
\end{IEEEproof}%

From \cite{Erez:Zamir:04}, we conclude the following lemma, which will be essential for our considerations in Section~\ref{sec:achievable_rates}.
\begin{lemma}[Erez$\cdot$Zamir'04]\label{lem:erez}
	There exists a sequence of nested lattices $\{\Lambda_{\textup{s}}^{(n)}\subset\Lambda_{\textup{c}}^{(n)}\}$ in which $\{\Lambda_{\textup{s}}^{(n)}\}$ is simultaneously good for AWGN channel coding and shaping and $\{\Lambda_{\textup{c}}^{(n)}\}$ for AWGN channel coding.  
\end{lemma}%
\begin{definition}\label{def:lattice_code}
	Given some pair of $n$-dimensional nested lattices $\Lambda_{\text{s}}\subset\Lambda_{\text{c}}$, a \emph{nested lattice code}
$\mathcal{C}^{(n)}$ is defined as
	\begin{equation}
		\mathcal{C}^{(n)}\coloneqq\Lambda_{\text{c}}\cap\mathcal{V}_{\text{s}}
		\label{eq:lattice_code}
	\end{equation}%
	with rate
	\begin{equation}
		R = \frac{1}{n}\log_2\bigl(\bigl|\mathcal{C}^{(n)}\bigr|\bigr)=\frac{1}{n}\log_2\left(\frac{\Vol(\mathcal{V}_{\text{s}})}{\Vol(\mathcal{V}_{\text{c}})}\right)\;.
		\label{eq:lattice_rate}
	\end{equation}%
\end{definition}%
\begin{remark}\label{rem:linearity}
	The essential structural property of a nested lattice code is linearity, which means that each sum of lattice codewords modulo
the shaping lattice is a codeword itself:
	\begin{equation}
		\ve{x}_1,\dots,\ve{x}_N\in\mathcal{C}^{(n)}\Rightarrow\left[\sum\nolimits_{i=1}^N\ve{x}_i\right]\bmod{\Lambda_{\text{s}}}\in\mathcal{C}^{(n)}\;.
		\label{eq:lattice_linearity}
	\end{equation}%
\end{remark}%
%
%
%
%--------------------------------------------------------------------------------------------------------------------------
\subsubsection{Channel Encoding}\label{sec:encoding}
In order to protect the modulo sum in (\ref{eq:mod_b_messages}) against Gaussian receiver noise, each sensor node employs the same $n$-dimensional nested lattice code $\mathcal{C}^{(n)}$ based on a nested lattice pair taken from Lemma~\ref{lem:erez}. The shaping lattice is scaled such that the second moment equals the transmit power constraint (i.e., $\sigma^2(\Lambda_{\text{s}})=P$). Thus, each node is equipped with the same \emph{channel encoder} (see Fig.~\ref{fig:nomographic})
\begin{equation}
	\mathcal{E}_2:\mathds{Z}_p^{k}\to\mathcal{C}^{(n)}\subset\mathds{R}^n\;,
	\label{eq:encoder}
\end{equation}%
which maps each message $\ve{w}_i$ to a length-$n$ lattice codeword:
\begin{equation*}
	\ve{x}_{i}=\bigl(x_i[1],\dots,x_i[n]\bigr)=\mathcal{E}_2(\ve{w}_{i})\;,\;i=1,\dots,N\;.
\end{equation*}%
Due to the scaling of the shaping lattice, each codeword meets the average power constraint and the message rate (\ref{eq:lattice_rate}) (in bits per channel use) is
\begin{equation}
	R = \frac{k}{n}\log_2(p)\;.
	\label{eq:message_rate}
\end{equation}%
In what follows, we assume that (\ref{eq:encoder}) fulfills
\begin{equation}
	\mathcal{E}_2^{-1}\left(\left[\sum\nolimits_{i=1}^N\mathcal{E}_2(\ve{w}_{i}\bigr)\right]\bmod{\Lambda_{\text{s}}}\right)=\bigoplus_{i=1}^N\ve{w}_i\;,
	\label{eq:bijectivity}
\end{equation}%
for all $\ve{w}_i\in\mathds{Z}_p^k$. The existence of such linearity preserving lattice encoders is shown in \cite[Lemma 6]{Nazer:Gastpar:11b}.\footnote{Note that $p$ is assumed to be prime in order to guarantee that $\mathds{Z}_p$ forms along with the addition and multiplication modulo $p$ a finite field, which is a necessary condition for the existence of nested lattice encoders fulfilling (\ref{eq:bijectivity}).}
%
%
% 
%--------------------------------------------------------------------------------------------------------------------------
\subsubsection{Channel Decoding}\label{sec:decoding}
After the Gaussian MAC has been used by the sensor nodes $n$ times, the FC is aware of the length-$n$ receive vector
\begin{equation}
	\ve{y}=\sum_{i=1}^N\ve{x}_{i}+\ve{z}\;,
	\label{eq:receive_vector}
\end{equation}%
where $\ve{z}\sim\mathcal{N}(\ve{0},\sigma_Z^2\ma{I}_n)$ (see Fig.~\ref{fig:nomographic}). In order to obtain an estimate of the modulo $p$ sum (\ref{eq:mod_b_messages}), the FC applies a channel decoder 
\begin{equation*}
	\mathcal{D}_2:\mathds{R}^n\to\mathds{Z}_p^{k}
\end{equation*}%
that consists of an \emph{Euclidean nearest neighbor decoder} \cite{Erez:Zamir:04} followed by the inverse of the channel encoding function. Thus, by (\ref{eq:bijectivity}) we have
\begin{equation}
	\hat{\ve{g}}=\mathcal{D}_2\bigl(\ve{y}\bigr)=\mathcal{E}_2^{-1}\left(\left[Q_{\Lambda_{\text{c}}}\bigl(\ve{y}\bigr)\right]\bmod{\Lambda_{\text{s}}}\right)\;.
	\label{eq:decoder}
\end{equation}%
Obviously, the nearest neighbor decoder quantizes the receive vector onto the coding lattice and then reduces the outcome to the shaping lattice in order to guarantee that the resulting lattice point is a valid codeword. 

Inserting (\ref{eq:receive_vector}) in (\ref{eq:decoder}) shows along with Lemma~\ref{lem:modulus} that
\setlength{\arraycolsep}{0.0em}
\begin{eqnarray}
	\hat{\ve{g}}&{}={}&\mathcal{E}_2^{-1}\left(\left[Q_{\Lambda_{\text{c}}}\left(\sum\nolimits_{i=1}^N\ve{x}_{i}+\ve{z}\right)\right]\bmod{\Lambda_{\text{s}}}\right)\nonumber\\
	&{}={}&\mathcal{E}_2^{-1}\Bigl(\Bigl[Q_{\Lambda_{\text{c}}}\Bigl(\Bigl[\sum\nolimits_{i=1}^N\ve{x}_{i}\Bigr]\bmod{\Lambda_{\text{s}}}+\ve{z}\Bigr)\Bigr]\bmod{\Lambda_{\text{s}}}\Bigr)\label{eq:decode2}\nonumber\\
	&{}={}&\mathcal{E}_2^{-1}\bigl(\left[Q_{\Lambda_{\text{c}}}\left(\ve{x}+\ve{z}\right)\right]\bmod{\Lambda_{\text{s}}}\bigr)\label{eq:inserted}\;,
\end{eqnarray}%
\setlength{\arraycolsep}{5pt}%
where $\ve{x}\coloneqq\left[\sum_{i}\ve{x}_{i}\right]\bmod{\Lambda_{\text{s}}}$. Because of (\ref{eq:lattice_linearity}), we have $\ve{x}\in\mathcal{C}^{(n)}$ so that (\ref{eq:receive_vector}) is essentially a single codeword corrupted by Gaussian noise. The computation over the channel can therefore be seen as a point-to-point link over which a single transmitter aims at reliably communicating the codeword $\ve{x}$ to its intended receiver.
%
%
%
%--------------------------------------------------------------------------------------------------------------------------
\subsubsection{Decoding Error Probability}\label{sec:error_probability}
Let $\delta>0$ be arbitrary. Then, the modulo $p$ sum of messages is said to be decoded with error probability $\delta$ if 
\begin{equation}
	P_e^{(n)}\coloneqq\Prob(\hat{\ve{g}}\neq\ve{g})\leq\delta\;.
	\label{eq:error_prob_def}
\end{equation}%
To demonstrate that this can be considered as a \emph{maximum probability of error}, we establish in the following the upper bound
\begin{equation}
	P_e^{(n)}\leq\Prob(\ve{z}\notin\mathcal{V}_{\text{c}})\;.
	\label{eq:error_prob_bound}
\end{equation}
Towards this end, note that each node chooses one out of $p^{k}$ codewords so that at the FC, the modulo $p$ sum (\ref{eq:mod_b_messages}) can take on at most $U\coloneqq\left(\begin{smallmatrix}p^k+N-1\\ N\end{smallmatrix}\right)$ different values $\ve{g}^{(u)}$, $u=1,\dots,U$. Thus, the conditional probability of error given that $\ve{g}^{(u)}$ is the correct sum leads with (\ref{eq:inserted}) to
\setlength{\arraycolsep}{0.0em}
\begin{eqnarray}
	\lambda^{(u)}&{}\coloneqq{}&\Prob\bigl(\hat{\ve{g}}\neq\ve{g}\,\big|\,\ve{g}=\ve{g}^{(u)}\bigr)\nonumber\\
	&\hspace{2pt}{}={}&\Prob\Bigl(\bigl[Q_{\Lambda_{\text{c}}}(\ve{x}+\ve{z})\bigr]\!\bmod{\Lambda_{\text{s}}}\neq\mathcal{E}_2(\ve{g})\,\Big|\,\ve{g}=\ve{g}^{(u)}\Bigr)\nonumber\\
	&\hspace{2pt}{}={}&\Prob(\ve{z}\notin\mathcal{V}_{\text{c}})\;.
	\label{eq:cond_error_prob}
\end{eqnarray}%
\setlength{\arraycolsep}{5pt}%
Observe that (\ref{eq:cond_error_prob}) is independent of $u$, which follows from the symmetry of the coding lattice $\Lambda_{\text{c}}$. Then, upper bounding the total probability as
\begin{equation}
	P_e^{(n)}=\sum_{u=1}^{U}\lambda^{(u)}\Prob\bigl(\ve{g}=\ve{g}^{(u)}\bigr)\leq\max_{1\leq u\leq U}\lambda^{(u)}=\Prob(\ve{z}\notin\mathcal{V}_{\text{c}})
	\label{eq:error_prob_sum}
\end{equation}%
shows that the decoding error probability is essentially a maximum probability of error.
\begin{remark}\label{rem:for_every_codeword}
	Note that in the network model given in Section~\ref{sec:network}, we did not introduce a probability distribution on the sensor readings, which requires the decoding error probability to be small for \emph{every} codeword and thus for every choice of $\{\varphi_{i}(s_i)\in\Pi\}_{i=1}^N$. According to (\ref{eq:error_prob_bound}), this can be ensured because if $\Prob(\ve{z}\notin\mathcal{V}_{\text{c}})\leq\delta$, then $P_e^{(n)}\leq\delta$, which justifies to consider (\ref{eq:error_prob_def}) as a maximum probability of error.
\end{remark}%
%
%
%
%%%%%%%%%%%%%%%%%%%%%%%%%%%%%%%%%%%%%%%%%%%%%%%%%%%%%%%%%%%%%%%%%%%%%%%%%%%%%%%%%%%%%%%%%%%%%%%%%%%%%%%%%%%%%%%%%%%%%%%%%%%
%%%%%%%%%%%%%%%%%%%%%%%%%%%%%%%%%%%%%%%%%%%%%%%%%%%%%%%%%%%%%%%%%%%%%%%%%%%%%%%%%%%%%%%%%%%%%%%%%%%%%%%%%%%%%%%%%%%%%%%%%%%
%	Achievable Computation Rates
%%%%%%%%%%%%%%%%%%%%%%%%%%%%%%%%%%%%%%%%%%%%%%%%%%%%%%%%%%%%%%%%%%%%%%%%%%%%%%%%%%%%%%%%%%%%%%%%%%%%%%%%%%%%%%%%%%%%%%%%%%%
%%%%%%%%%%%%%%%%%%%%%%%%%%%%%%%%%%%%%%%%%%%%%%%%%%%%%%%%%%%%%%%%%%%%%%%%%%%%%%%%%%%%%%%%%%%%%%%%%%%%%%%%%%%%%%%%%%%%%%%%%%%
%
\section{Achievable Computation Rates} \label{sec:achievable_rates}
In this section, our objective is to characterize the computation rates that are achievable in clustered Gaussian sensor networks with the scheme of Section~\ref{sec:reliable_computation}. In order to gain first insights, we start in Section~\ref{sec:single_cluster} with a single cluster network followed by the general case in Section~\ref{sec:multiple_cluster}. Section~\ref{sec:discussion} is then devoted to discuss the results. First, however, we provide a lemma that guarantees that with the quantization procedure of Section~\ref{sec:processing}, the desired functions in (\ref{eq:kolmogorov_cluster_alt}) can be approximated with arbitrary precision.
\begin{lemma}\label{lem:approximation_error}
	Let $(f_1,\dots,f_L)\in\C^0(\mathds{E}^{|C_1|})\times\cdots\times\C^0(\mathds{E}^{|C_L|})$ be some choice of $L$ Kolmogorov's superpositions. Then, each $f_{\ell}$ can be uniformly approximated with arbitrary precision $\varepsilon>0$ if the common quantizer (\ref{eq:quantizer}) is configured with sufficiently large $b=b(f_1,\dots,f_L,N,\varepsilon)$. That is,
\begin{equation*}
	\forall\varepsilon>0\,\exists b_0\,\forall\ell\,\forall b\geq b_0:\sup_{\ve{s}_{\ell}\in\mathds{E}^{|C_{\ell}|}}\left|f_{\ell}(\ve{s}_{\ell})-\tilde{f}_{\ell}(\ve{s}_{\ell})\right|<\varepsilon\;.
\end{equation*}%
\end{lemma}%
\begin{IEEEproof}
	The proof is deferred to Appendix~\ref{app:lemma3}.
\end{IEEEproof}%
\begin{remark}\label{rem:N_dependency}
	In words, the quantization parameter $b_0=b_0(f_1,\dots,f_L,N,\varepsilon)$ denotes the smallest number of bits with which $f_1,\dots,f_L$ can be represented within accuracy $\varepsilon$. We point out, however, that it is not a particular property of the scheme presented in Section~\ref{sec:reliable_computation} that $b_0$ generally depends also on the number of nodes. In fact, all computation schemes that approximate a multivariate function by quantizing its arguments suffer from this. Hence, we drop the corresponding indication in what follows.   
\end{remark}
\begin{remark}\label{rem:max_prob}
	Due to Remark \ref{rem:for_every_codeword}, (\ref{eq:computation_rate_error}) represents a maximum error probability, which is therefore independent of the statistics of sensor readings. Because of Lemma~\ref{lem:approximation_error}, it can therefore be written as $\Prob(\bigcup_{t=1}^T\{\hat{f}(\ve{s}[t])\neq\tilde{f}(\ve{s}[t])\})$, with $\tilde{f}$ as defined in~(\ref{eq:f_approx}).
\end{remark}%
%
%
%
%
%%%%%%%%%%%%%%%%%%%%%%%%%%%%%%%%%%%%%%%%%%%%%%%%%%%%%%%%%%%%%%%%%%%%%%%%%%%%%%%%%%%%%%%%%%%%%%%%%%%%%%%%%%%%%%%%%%%%%%%%%%%
\subsection{The Single Cluster Case} \label{sec:single_cluster}
Consider a single cluster consisting of $N$ nodes, which means that $L=1$ and $|C_1|=N$. 
%
%%%%%%%%%%%%%%%%%%%%%%%%%%%%%%%%%%%%%%%%%%%%%%%%%%%%%%%%%%%%%%%%%%%%%%%%%%%%%%%%%%%%%%%%%%%%%%%%%%%%%%%%%%%%%%%%%%%%%%%%%%%
\subsubsection{Nomographic Functions} \label{sec:nomographic_single}
We start with the computation of a single nomographic function over a Gaussian MAC such as depicted in Fig.~\ref{fig:nomographic}. The following theorem provides an achievable rate at which elements from $\Nomo^0(\mathds{E}^N)$ can be reliably computed through harnessing the interference. Note that according to (\ref{eq:decoder_1}), the estimate $\hat{f}$ of some given $f\in\Nomo^0(\mathds{E}^N)$ is defined to be (see Fig.~\ref{fig:nomographic})
\begin{equation}
	\hat{f}\bigl(\ve{s}[t]\bigr)=\psi\left(Q^{-1}(\hat{\ve{g}}[t])\right)\;,\;t=1,\dots,T\;.
	\label{eq:nomo_estimate}
\end{equation}%
\begin{theorem}\label{thm:sum_protection}
	Given $f\in\Nomo^0(\mathds{E}^N)$, let $\hat{f}$ be its estimate defined by (\ref{eq:nomo_estimate}). Let $\varepsilon>0$ be some given desired accuracy and $b_0(f,\varepsilon)$ be specified as in Lemma~\ref{lem:approximation_error}. Then,
	\begin{equation}
		R^{\textup{Comp}}(f,\varepsilon)=\frac{\frac{1}{2}\log_2^+\left(\frac{P}{\sigma_Z^2}\right)}{b_0(f,\varepsilon)+\log_2(N)}
		\label{eq:max_rate}
	\end{equation}%
is an achievable computation rate for $f$ and $\varepsilon$.
\end{theorem}%
\begin{IEEEproof}
	The proof is deferred to Appendix~\ref{app:theorem5}.
\end{IEEEproof}%
\begin{remark}\label{rem:bit_more}
	Note that in accordance with the proof of the theorem, (\ref{eq:max_rate}) can even be slightly improved if the bound in (\ref{eq:tau_bound}) is applied instead of (\ref{eq:tau_bound2}).
\end{remark}%

In the following, we present some examples that are reliably computable up to rate (\ref{eq:max_rate}).
\begin{example}[Arithmetic Mean]\label{ex:arithmetic_mean}
	Let the desired function be the arithmetic mean $f(s_1,\dots,s_N)=\frac{1}{N}\sum_{i=1}^Ns_i$. With the continuous pre-processing functions $\varphi_i(s_i)=s_i$, $i=1,\dots,N$, and the continuous post-processing function $\psi(g)=\frac{1}{N}g$, we have $f\in\Nomo^0(\mathds{E}^N)$.
\end{example}%

\begin{example}[Geometric Mean]\label{ex:geometric_mean}
	Let the desired function be the geometric mean $f(s_1,\dots,s_N)=\bigl(\prod_{i=1}^Ns_i\bigr)^{1/N}$ with domain $[s_{\text{min}},1]^N$, for some $0< s_{\text{min}}<1$.
With the continuous pre-processing functions $\varphi_i(s_i)=\log_{\e}(s_i)$, $i=1,\dots,N$, and the continuous post-processing function $\psi(g)=\exp_{\e}(g/N)$, we have
$f\in\Nomo^0([s_{\text{min}},1]^N)$.
\end{example}%
\begin{example}[Euclidean Norm]\label{ex:Euclidean_norm}
	Let the desired function be the Euclidean norm $f(s_1,\dots,s_K)=\sqrt{s_1^2+\cdots +s_N^2}$. With the continuous pre-processing functions $\varphi_i(s_i)=s_i^2$, $i=1,\dots,N$, and the continuous post-processing function $\psi(g)=\sqrt{g}$, we have $f\in\Nomo^0(\mathds{E}^N)$.
\end{example}%
\begin{figure}[!t]
	\centering
	\begin{tikzpicture}
		\pgfplotsset{every axis legend/.append style={at={(0.02,0.97)},font=\small, anchor=north west, cells={anchor=west}}}
		\pgfplotsset{every axis plot/.append style={thick}}
		\pgfsetplotmarkrepeat{50}
		\begin{axis}[
			width=0.47\textwidth,
			height=0.335\textwidth,
			xmin=0, xmax=20,
			ymin=0, ymax=0.25,
			yticklabel style={/pgf/number format/fixed},
			enlargelimits=false,
			xlabel={$P/\sigma_Z^2$ in dB},
			ylabel={Computation Rate},
			xmajorgrids,
			ymajorgrids,
			legend entries={Arithmetic Mean, Geometric Mean, Euclidean Norm}
			]

			\addplot [
			color=black,
			solid,
			mark=star,
			mark options={scale=1.5,solid,fill=white,thick}
			]coordinates{
			 (0,0) (0.1,0.00124679) (0.2,0.00249358) (0.3,0.00374037) (0.4,0.00498716) (0.5,0.00623395) (0.6,0.00748074) (0.7,0.00872753) (0.8,0.00997432) (0.9,0.0112211) (1,0.0124679) (1.1,0.0137147) (1.2,0.0149615) (1.3,0.0162083) (1.4,0.0174551) (1.5,0.0187018) (1.6,0.0199486) (1.7,0.0211954) (1.8,0.0224422) (1.9,0.023689) (2,0.0249358) (2.1,0.0261826) (2.2,0.0274294) (2.3,0.0286762) (2.4,0.0299229) (2.5,0.0311697) (2.6,0.0324165) (2.7,0.0336633) (2.8,0.0349101) (2.9,0.0361569) (3,0.0374037) (3.1,0.0386505) (3.2,0.0398973) (3.3,0.0411441) (3.4,0.0423908) (3.5,0.0436376) (3.6,0.0448844) (3.7,0.0461312) (3.8,0.047378) (3.9,0.0486248) (4,0.0498716) (4.1,0.0511184) (4.2,0.0523652) (4.3,0.0536119) (4.4,0.0548587) (4.5,0.0561055) (4.6,0.0573523) (4.7,0.0585991) (4.8,0.0598459) (4.9,0.0610927) (5,0.0623395) (5.1,0.0635863) (5.2,0.0648331) (5.3,0.0660798) (5.4,0.0673266) (5.5,0.0685734) (5.6,0.0698202) (5.7,0.071067) (5.8,0.0723138) (5.9,0.0735606) (6,0.0748074) (6.1,0.0760542) (6.2,0.077301) (6.3,0.0785477) (6.4,0.0797945) (6.5,0.0810413) (6.6,0.0822881) (6.7,0.0835349) (6.8,0.0847817) (6.9,0.0860285) (7,0.0872753) (7.1,0.0885221) (7.2,0.0897688) (7.3,0.0910156) (7.4,0.0922624) (7.5,0.0935092) (7.6,0.094756) (7.7,0.0960028) (7.8,0.0972496) (7.9,0.0984964) (8,0.0997432) (8.1,0.10099) (8.2,0.102237) (8.3,0.103484) (8.4,0.10473) (8.5,0.105977) (8.6,0.107224) (8.7,0.108471) (8.8,0.109717) (8.9,0.110964) (9,0.112211) (9.1,0.113458) (9.2,0.114705) (9.3,0.115951) (9.4,0.117198) (9.5,0.118445) (9.6,0.119692) (9.7,0.120939) (9.8,0.122185) (9.9,0.123432) (10,0.124679) (10.1,0.125926) (10.2,0.127173) (10.3,0.128419) (10.4,0.129666) (10.5,0.130913) (10.6,0.13216) (10.7,0.133406) (10.8,0.134653) (10.9,0.1359) (11,0.137147) (11.1,0.138394) (11.2,0.13964) (11.3,0.140887) (11.4,0.142134) (11.5,0.143381) (11.6,0.144628) (11.7,0.145874) (11.8,0.147121) (11.9,0.148368) (12,0.149615) (12.1,0.150862) (12.2,0.152108) (12.3,0.153355) (12.4,0.154602) (12.5,0.155849) (12.6,0.157095) (12.7,0.158342) (12.8,0.159589) (12.9,0.160836) (13,0.162083) (13.1,0.163329) (13.2,0.164576) (13.3,0.165823) (13.4,0.16707) (13.5,0.168317) (13.6,0.169563) (13.7,0.17081) (13.8,0.172057) (13.9,0.173304) (14,0.174551) (14.1,0.175797) (14.2,0.177044) (14.3,0.178291) (14.4,0.179538) (14.5,0.180784) (14.6,0.182031) (14.7,0.183278) (14.8,0.184525) (14.9,0.185772) (15,0.187018) (15.1,0.188265) (15.2,0.189512) (15.3,0.190759) (15.4,0.192006) (15.5,0.193252) (15.6,0.194499) (15.7,0.195746) (15.8,0.196993) (15.9,0.19824) (16,0.199486) (16.1,0.200733) (16.2,0.20198) (16.3,0.203227) (16.4,0.204473) (16.5,0.20572) (16.6,0.206967) (16.7,0.208214) (16.8,0.209461) (16.9,0.210707) (17,0.211954) (17.1,0.213201) (17.2,0.214448) (17.3,0.215695) (17.4,0.216941) (17.5,0.218188) (17.6,0.219435) (17.7,0.220682) (17.8,0.221929) (17.9,0.223175) (18,0.224422) (18.1,0.225669) (18.2,0.226916) (18.3,0.228162) (18.4,0.229409) (18.5,0.230656) (18.6,0.231903) (18.7,0.23315) (18.8,0.234396) (18.9,0.235643) (19,0.23689) (19.1,0.238137) (19.2,0.239384) (19.3,0.24063) (19.4,0.241877) (19.5,0.243124) (19.6,0.244371) (19.7,0.245618) (19.8,0.246864) (19.9,0.248111) (20,0.249358)
			};

			\addplot [
			color=black,
			dashed
			]
			plot[forget plot]
			coordinates{
			 (0,0.0375321) (0.1,0.0381591) (0.2,0.0387932) (0.3,0.0394346) (0.4,0.0400831) (0.5,0.0407387) (0.6,0.0414016) (0.7,0.0420715) (0.8,0.0427486) (0.9,0.0434328) (1,0.0441241) (1.1,0.0448225) (1.2,0.045528) (1.3,0.0462405) (1.4,0.04696) (1.5,0.0476865) (1.6,0.0484199) (1.7,0.0491604) (1.8,0.0499077) (1.9,0.0506619) (2,0.0514229) (2.1,0.0521908) (2.2,0.0529654) (2.3,0.0537467) (2.4,0.0545348) (2.5,0.0553295) (2.6,0.0561308) (2.7,0.0569387) (2.8,0.0577532) (2.9,0.0585741) (3,0.0594014) (3.1,0.0602351) (3.2,0.0610752) (3.3,0.0619215) (3.4,0.0627741) (3.5,0.0636328) (3.6,0.0644977) (3.7,0.0653687) (3.8,0.0662456) (3.9,0.0671286) (4,0.0680174) (4.1,0.0689121) (4.2,0.0698126) (4.3,0.0707188) (4.4,0.0716306) (4.5,0.0725481) (4.6,0.0734712) (4.7,0.0743997) (4.8,0.0753336) (4.9,0.0762729) (5,0.0772175) (5.1,0.0781674) (5.2,0.0791224) (5.3,0.0800826) (5.4,0.0810478) (5.5,0.082018) (5.6,0.0829931) (5.7,0.083973) (5.8,0.0849578) (5.9,0.0859473) (6,0.0869415) (6.1,0.0879403) (6.2,0.0889436) (6.3,0.0899514) (6.4,0.0909636) (6.5,0.0919802) (6.6,0.093001) (6.7,0.0940261) (6.8,0.0950554) (6.9,0.0960887) (7,0.0971261) (7.1,0.0981675) (7.2,0.0992128) (7.3,0.100262) (7.4,0.101315) (7.5,0.102372) (7.6,0.103432) (7.7,0.104496) (7.8,0.105564) (7.9,0.106635) (8,0.107709) (8.1,0.108787) (8.2,0.109868) (8.3,0.110953) (8.4,0.112041) (8.5,0.113131) (8.6,0.114225) (8.7,0.115323) (8.8,0.116423) (8.9,0.117526) (9,0.118632) (9.1,0.11974) (9.2,0.120852) (9.3,0.121966) (9.4,0.123083) (9.5,0.124203) (9.6,0.125325) (9.7,0.12645) (9.8,0.127578) (9.9,0.128707) (10,0.12984) (10.1,0.130974) (10.2,0.132111) (10.3,0.133251) (10.4,0.134392) (10.5,0.135536) (10.6,0.136682) (10.7,0.13783) (10.8,0.13898) (10.9,0.140132) (11,0.141286) (11.1,0.142442) (11.2,0.1436) (11.3,0.144759) (11.4,0.145921) (11.5,0.147085) (11.6,0.14825) (11.7,0.149417) (11.8,0.150585) (11.9,0.151756) (12,0.152928) (12.1,0.154101) (12.2,0.155277) (12.3,0.156453) (12.4,0.157631) (12.5,0.158811) (12.6,0.159992) (12.7,0.161175) (12.8,0.162359) (12.9,0.163544) (13,0.164731) (13.1,0.165919) (13.2,0.167108) (13.3,0.168298) (13.4,0.16949) (13.5,0.170683) (13.6,0.171877) (13.7,0.173072) (13.8,0.174268) (13.9,0.175466) (14,0.176664) (14.1,0.177864) (14.2,0.179065) (14.3,0.180266) (14.4,0.181469) (14.5,0.182672) (14.6,0.183877) (14.7,0.185082) (14.8,0.186289) (14.9,0.187496) (15,0.188704) (15.1,0.189913) (15.2,0.191123) (15.3,0.192334) (15.4,0.193545) (15.5,0.194757) (15.6,0.19597) (15.7,0.197184) (15.8,0.198399) (15.9,0.199614) (16,0.20083) (16.1,0.202046) (16.2,0.203263) (16.3,0.204481) (16.4,0.2057) (16.5,0.206919) (16.6,0.208139) (16.7,0.209359) (16.8,0.21058) (16.9,0.211802) (17,0.213024) (17.1,0.214247) (17.2,0.21547) (17.3,0.216694) (17.4,0.217918) (17.5,0.219143) (17.6,0.220368) (17.7,0.221594) (17.8,0.22282) (17.9,0.224046) (18,0.225274) (18.1,0.226501) (18.2,0.227729) (18.3,0.228958) (18.4,0.230186) (18.5,0.231416) (18.6,0.232645) (18.7,0.233875) (18.8,0.235106) (18.9,0.236336) (19,0.237567) (19.1,0.238799) (19.2,0.240031) (19.3,0.241263) (19.4,0.242495) (19.5,0.243728) (19.6,0.244961) (19.7,0.246195) (19.8,0.247428) (19.9,0.248662) (20,0.249897)
			};

			\addplot [
			color=black,
			solid,
			mark=square*,
			mark options={solid,fill=white,thick}
			]coordinates{
			 (0,0) (0.1,0.000958879) (0.2,0.00191776) (0.3,0.00287664) (0.4,0.00383552) (0.5,0.0047944) (0.6,0.00575328) (0.7,0.00671216) (0.8,0.00767104) (0.9,0.00862991) (1,0.00958879) (1.1,0.0105477) (1.2,0.0115066) (1.3,0.0124654) (1.4,0.0134243) (1.5,0.0143832) (1.6,0.0153421) (1.7,0.016301) (1.8,0.0172598) (1.9,0.0182187) (2,0.0191776) (2.1,0.0201365) (2.2,0.0210953) (2.3,0.0220542) (2.4,0.0230131) (2.5,0.023972) (2.6,0.0249309) (2.7,0.0258897) (2.8,0.0268486) (2.9,0.0278075) (3,0.0287664) (3.1,0.0297253) (3.2,0.0306841) (3.3,0.031643) (3.4,0.0326019) (3.5,0.0335608) (3.6,0.0345197) (3.7,0.0354785) (3.8,0.0364374) (3.9,0.0373963) (4,0.0383552) (4.1,0.0393141) (4.2,0.0402729) (4.3,0.0412318) (4.4,0.0421907) (4.5,0.0431496) (4.6,0.0441085) (4.7,0.0450673) (4.8,0.0460262) (4.9,0.0469851) (5,0.047944) (5.1,0.0489029) (5.2,0.0498617) (5.3,0.0508206) (5.4,0.0517795) (5.5,0.0527384) (5.6,0.0536972) (5.7,0.0546561) (5.8,0.055615) (5.9,0.0565739) (6,0.0575328) (6.1,0.0584916) (6.2,0.0594505) (6.3,0.0604094) (6.4,0.0613683) (6.5,0.0623272) (6.6,0.063286) (6.7,0.0642449) (6.8,0.0652038) (6.9,0.0661627) (7,0.0671216) (7.1,0.0680804) (7.2,0.0690393) (7.3,0.0699982) (7.4,0.0709571) (7.5,0.071916) (7.6,0.0728748) (7.7,0.0738337) (7.8,0.0747926) (7.9,0.0757515) (8,0.0767104) (8.1,0.0776692) (8.2,0.0786281) (8.3,0.079587) (8.4,0.0805459) (8.5,0.0815048) (8.6,0.0824636) (8.7,0.0834225) (8.8,0.0843814) (8.9,0.0853403) (9,0.0862991) (9.1,0.087258) (9.2,0.0882169) (9.3,0.0891758) (9.4,0.0901347) (9.5,0.0910935) (9.6,0.0920524) (9.7,0.0930113) (9.8,0.0939702) (9.9,0.0949291) (10,0.0958879) (10.1,0.0968468) (10.2,0.0978057) (10.3,0.0987646) (10.4,0.0997235) (10.5,0.100682) (10.6,0.101641) (10.7,0.1026) (10.8,0.103559) (10.9,0.104518) (11,0.105477) (11.1,0.106436) (11.2,0.107394) (11.3,0.108353) (11.4,0.109312) (11.5,0.110271) (11.6,0.11123) (11.7,0.112189) (11.8,0.113148) (11.9,0.114107) (12,0.115066) (12.1,0.116024) (12.2,0.116983) (12.3,0.117942) (12.4,0.118901) (12.5,0.11986) (12.6,0.120819) (12.7,0.121778) (12.8,0.122737) (12.9,0.123695) (13,0.124654) (13.1,0.125613) (13.2,0.126572) (13.3,0.127531) (13.4,0.12849) (13.5,0.129449) (13.6,0.130408) (13.7,0.131366) (13.8,0.132325) (13.9,0.133284) (14,0.134243) (14.1,0.135202) (14.2,0.136161) (14.3,0.13712) (14.4,0.138079) (14.5,0.139038) (14.6,0.139996) (14.7,0.140955) (14.8,0.141914) (14.9,0.142873) (15,0.143832) (15.1,0.144791) (15.2,0.14575) (15.3,0.146709) (15.4,0.147667) (15.5,0.148626) (15.6,0.149585) (15.7,0.150544) (15.8,0.151503) (15.9,0.152462) (16,0.153421) (16.1,0.15438) (16.2,0.155338) (16.3,0.156297) (16.4,0.157256) (16.5,0.158215) (16.6,0.159174) (16.7,0.160133) (16.8,0.161092) (16.9,0.162051) (17,0.16301) (17.1,0.163968) (17.2,0.164927) (17.3,0.165886) (17.4,0.166845) (17.5,0.167804) (17.6,0.168763) (17.7,0.169722) (17.8,0.170681) (17.9,0.171639) (18,0.172598) (18.1,0.173557) (18.2,0.174516) (18.3,0.175475) (18.4,0.176434) (18.5,0.177393) (18.6,0.178352) (18.7,0.17931) (18.8,0.180269) (18.9,0.181228) (19,0.182187) (19.1,0.183146) (19.2,0.184105) (19.3,0.185064) (19.4,0.186023) (19.5,0.186981) (19.6,0.18794) (19.7,0.188899) (19.8,0.189858) (19.9,0.190817) (20,0.191776)
			};

			\addplot [
			color=black,
			dashed
			]
			plot[forget plot]
			coordinates{
			 (0,0.0288651) (0.1,0.0293473) (0.2,0.0298351) (0.3,0.0303283) (0.4,0.030827) (0.5,0.0313313) (0.6,0.0318411) (0.7,0.0323563) (0.8,0.032877) (0.9,0.0334033) (1,0.0339349) (1.1,0.034472) (1.2,0.0350146) (1.3,0.0355626) (1.4,0.0361159) (1.5,0.0366747) (1.6,0.0372388) (1.7,0.0378082) (1.8,0.0383829) (1.9,0.038963) (2,0.0395483) (2.1,0.0401388) (2.2,0.0407346) (2.3,0.0413355) (2.4,0.0419416) (2.5,0.0425528) (2.6,0.043169) (2.7,0.0437904) (2.8,0.0444167) (2.9,0.0450481) (3,0.0456844) (3.1,0.0463256) (3.2,0.0469716) (3.3,0.0476225) (3.4,0.0482782) (3.5,0.0489387) (3.6,0.0496038) (3.7,0.0502737) (3.8,0.0509481) (3.9,0.0516272) (4,0.0523108) (4.1,0.0529988) (4.2,0.0536914) (4.3,0.0543883) (4.4,0.0550896) (4.5,0.0557952) (4.6,0.0565051) (4.7,0.0572192) (4.8,0.0579375) (4.9,0.0586599) (5,0.0593864) (5.1,0.0601169) (5.2,0.0608514) (5.3,0.0615898) (5.4,0.0623321) (5.5,0.0630783) (5.6,0.0638282) (5.7,0.0645819) (5.8,0.0653392) (5.9,0.0661003) (6,0.0668648) (6.1,0.067633) (6.2,0.0684046) (6.3,0.0691797) (6.4,0.0699582) (6.5,0.07074) (6.6,0.0715251) (6.7,0.0723135) (6.8,0.0731051) (6.9,0.0738998) (7,0.0746976) (7.1,0.0754985) (7.2,0.0763025) (7.3,0.0771094) (7.4,0.0779192) (7.5,0.0787318) (7.6,0.0795474) (7.7,0.0803657) (7.8,0.0811867) (7.9,0.0820104) (8,0.0828368) (8.1,0.0836658) (8.2,0.0844974) (8.3,0.0853315) (8.4,0.086168) (8.5,0.087007) (8.6,0.0878484) (8.7,0.0886921) (8.8,0.0895382) (8.9,0.0903865) (9,0.0912371) (9.1,0.0920898) (9.2,0.0929447) (9.3,0.0938018) (9.4,0.0946609) (9.5,0.095522) (9.6,0.0963851) (9.7,0.0972502) (9.8,0.0981173) (9.9,0.0989862) (10,0.099857) (10.1,0.10073) (10.2,0.101604) (10.3,0.10248) (10.4,0.103358) (10.5,0.104238) (10.6,0.105119) (10.7,0.106002) (10.8,0.106886) (10.9,0.107772) (11,0.10866) (11.1,0.109549) (11.2,0.110439) (11.3,0.111331) (11.4,0.112225) (11.5,0.11312) (11.6,0.114016) (11.7,0.114913) (11.8,0.115812) (11.9,0.116712) (12,0.117614) (12.1,0.118516) (12.2,0.11942) (12.3,0.120325) (12.4,0.121231) (12.5,0.122138) (12.6,0.123047) (12.7,0.123956) (12.8,0.124867) (12.9,0.125778) (13,0.126691) (13.1,0.127604) (13.2,0.128519) (13.3,0.129435) (13.4,0.130351) (13.5,0.131269) (13.6,0.132187) (13.7,0.133106) (13.8,0.134026) (13.9,0.134947) (14,0.135869) (14.1,0.136791) (14.2,0.137715) (14.3,0.138639) (14.4,0.139564) (14.5,0.140489) (14.6,0.141416) (14.7,0.142343) (14.8,0.143271) (14.9,0.144199) (15,0.145128) (15.1,0.146058) (15.2,0.146989) (15.3,0.14792) (15.4,0.148851) (15.5,0.149784) (15.6,0.150717) (15.7,0.15165) (15.8,0.152584) (15.9,0.153519) (16,0.154454) (16.1,0.155389) (16.2,0.156326) (16.3,0.157262) (16.4,0.158199) (16.5,0.159137) (16.6,0.160075) (16.7,0.161014) (16.8,0.161953) (16.9,0.162892) (17,0.163832) (17.1,0.164773) (17.2,0.165713) (17.3,0.166654) (17.4,0.167596) (17.5,0.168538) (17.6,0.16948) (17.7,0.170423) (17.8,0.171366) (17.9,0.172309) (18,0.173253) (18.1,0.174197) (18.2,0.175142) (18.3,0.176086) (18.4,0.177031) (18.5,0.177977) (18.6,0.178922) (18.7,0.179868) (18.8,0.180815) (18.9,0.181761) (19,0.182708) (19.1,0.183655) (19.2,0.184603) (19.3,0.18555) (19.4,0.186498) (19.5,0.187446) (19.6,0.188394) (19.7,0.189343) (19.8,0.190292) (19.9,0.191241) (20,0.19219)
			};

			\addplot [
			color=black,
			solid,
			mark=diamond*,
			mark options={scale=1.5,solid,fill=white,thick}
			]coordinates{
			 (0,0) (0.1,0.000631019) (0.2,0.00126204) (0.3,0.00189306) (0.4,0.00252408) (0.5,0.0031551) (0.6,0.00378611) (0.7,0.00441713) (0.8,0.00504815) (0.9,0.00567917) (1,0.00631019) (1.1,0.00694121) (1.2,0.00757223) (1.3,0.00820325) (1.4,0.00883427) (1.5,0.00946529) (1.6,0.0100963) (1.7,0.0107273) (1.8,0.0113583) (1.9,0.0119894) (2,0.0126204) (2.1,0.0132514) (2.2,0.0138824) (2.3,0.0145134) (2.4,0.0151445) (2.5,0.0157755) (2.6,0.0164065) (2.7,0.0170375) (2.8,0.0176685) (2.9,0.0182996) (3,0.0189306) (3.1,0.0195616) (3.2,0.0201926) (3.3,0.0208236) (3.4,0.0214547) (3.5,0.0220857) (3.6,0.0227167) (3.7,0.0233477) (3.8,0.0239787) (3.9,0.0246097) (4,0.0252408) (4.1,0.0258718) (4.2,0.0265028) (4.3,0.0271338) (4.4,0.0277648) (4.5,0.0283959) (4.6,0.0290269) (4.7,0.0296579) (4.8,0.0302889) (4.9,0.0309199) (5,0.031551) (5.1,0.032182) (5.2,0.032813) (5.3,0.033444) (5.4,0.034075) (5.5,0.0347061) (5.6,0.0353371) (5.7,0.0359681) (5.8,0.0365991) (5.9,0.0372301) (6,0.0378611) (6.1,0.0384922) (6.2,0.0391232) (6.3,0.0397542) (6.4,0.0403852) (6.5,0.0410162) (6.6,0.0416473) (6.7,0.0422783) (6.8,0.0429093) (6.9,0.0435403) (7,0.0441713) (7.1,0.0448024) (7.2,0.0454334) (7.3,0.0460644) (7.4,0.0466954) (7.5,0.0473264) (7.6,0.0479575) (7.7,0.0485885) (7.8,0.0492195) (7.9,0.0498505) (8,0.0504815) (8.1,0.0511126) (8.2,0.0517436) (8.3,0.0523746) (8.4,0.0530056) (8.5,0.0536366) (8.6,0.0542676) (8.7,0.0548987) (8.8,0.0555297) (8.9,0.0561607) (9,0.0567917) (9.1,0.0574227) (9.2,0.0580538) (9.3,0.0586848) (9.4,0.0593158) (9.5,0.0599468) (9.6,0.0605778) (9.7,0.0612089) (9.8,0.0618399) (9.9,0.0624709) (10,0.0631019) (10.1,0.0637329) (10.2,0.064364) (10.3,0.064995) (10.4,0.065626) (10.5,0.066257) (10.6,0.066888) (10.7,0.067519) (10.8,0.0681501) (10.9,0.0687811) (11,0.0694121) (11.1,0.0700431) (11.2,0.0706741) (11.3,0.0713052) (11.4,0.0719362) (11.5,0.0725672) (11.6,0.0731982) (11.7,0.0738292) (11.8,0.0744603) (11.9,0.0750913) (12,0.0757223) (12.1,0.0763533) (12.2,0.0769843) (12.3,0.0776154) (12.4,0.0782464) (12.5,0.0788774) (12.6,0.0795084) (12.7,0.0801394) (12.8,0.0807705) (12.9,0.0814015) (13,0.0820325) (13.1,0.0826635) (13.2,0.0832945) (13.3,0.0839255) (13.4,0.0845566) (13.5,0.0851876) (13.6,0.0858186) (13.7,0.0864496) (13.8,0.0870806) (13.9,0.0877117) (14,0.0883427) (14.1,0.0889737) (14.2,0.0896047) (14.3,0.0902357) (14.4,0.0908668) (14.5,0.0914978) (14.6,0.0921288) (14.7,0.0927598) (14.8,0.0933908) (14.9,0.0940219) (15,0.0946529) (15.1,0.0952839) (15.2,0.0959149) (15.3,0.0965459) (15.4,0.0971769) (15.5,0.097808) (15.6,0.098439) (15.7,0.09907) (15.8,0.099701) (15.9,0.100332) (16,0.100963) (16.1,0.101594) (16.2,0.102225) (16.3,0.102856) (16.4,0.103487) (16.5,0.104118) (16.6,0.104749) (16.7,0.10538) (16.8,0.106011) (16.9,0.106642) (17,0.107273) (17.1,0.107904) (17.2,0.108535) (17.3,0.109166) (17.4,0.109797) (17.5,0.110428) (17.6,0.111059) (17.7,0.11169) (17.8,0.112321) (17.9,0.112952) (18,0.113583) (18.1,0.114214) (18.2,0.114845) (18.3,0.115477) (18.4,0.116108) (18.5,0.116739) (18.6,0.11737) (18.7,0.118001) (18.8,0.118632) (18.9,0.119263) (19,0.119894) (19.1,0.120525) (19.2,0.121156) (19.3,0.121787) (19.4,0.122418) (19.5,0.123049) (19.6,0.12368) (19.7,0.124311) (19.8,0.124942) (19.9,0.125573) (20,0.126204)
			};

			\addplot [
			color=black,
			dashed
			]
			plot[forget plot]
			coordinates{
			 (0,0.0189956) (0.1,0.0193129) (0.2,0.0196339) (0.3,0.0199584) (0.4,0.0202867) (0.5,0.0206185) (0.6,0.020954) (0.7,0.021293) (0.8,0.0216357) (0.9,0.021982) (1,0.0223319) (1.1,0.0226854) (1.2,0.0230424) (1.3,0.023403) (1.4,0.0237672) (1.5,0.0241348) (1.6,0.0245061) (1.7,0.0248808) (1.8,0.025259) (1.9,0.0256407) (2,0.0260259) (2.1,0.0264145) (2.2,0.0268066) (2.3,0.027202) (2.4,0.0276009) (2.5,0.0280031) (2.6,0.0284087) (2.7,0.0288176) (2.8,0.0292298) (2.9,0.0296452) (3,0.030064) (3.1,0.0304859) (3.2,0.0309111) (3.3,0.0313394) (3.4,0.0317709) (3.5,0.0322055) (3.6,0.0326433) (3.7,0.0330841) (3.8,0.0335279) (3.9,0.0339748) (4,0.0344246) (4.1,0.0348775) (4.2,0.0353332) (4.3,0.0357918) (4.4,0.0362534) (4.5,0.0367177) (4.6,0.0371849) (4.7,0.0376548) (4.8,0.0381275) (4.9,0.0386029) (5,0.039081) (5.1,0.0395617) (5.2,0.0400451) (5.3,0.040531) (5.4,0.0410195) (5.5,0.0415105) (5.6,0.0420041) (5.7,0.0425) (5.8,0.0429984) (5.9,0.0434992) (6,0.0440024) (6.1,0.0445079) (6.2,0.0450157) (6.3,0.0455258) (6.4,0.0460381) (6.5,0.0465526) (6.6,0.0470692) (6.7,0.047588) (6.8,0.048109) (6.9,0.048632) (7,0.049157) (7.1,0.0496841) (7.2,0.0502131) (7.3,0.0507441) (7.4,0.051277) (7.5,0.0518118) (7.6,0.0523485) (7.7,0.052887) (7.8,0.0534273) (7.9,0.0539694) (8,0.0545132) (8.1,0.0550588) (8.2,0.055606) (8.3,0.0561549) (8.4,0.0567054) (8.5,0.0572576) (8.6,0.0578113) (8.7,0.0583665) (8.8,0.0589233) (8.9,0.0594815) (9,0.0600413) (9.1,0.0606025) (9.2,0.0611651) (9.3,0.061729) (9.4,0.0622944) (9.5,0.0628611) (9.6,0.0634291) (9.7,0.0639984) (9.8,0.064569) (9.9,0.0651408) (10,0.0657139) (10.1,0.0662881) (10.2,0.0668636) (10.3,0.0674401) (10.4,0.0680179) (10.5,0.0685967) (10.6,0.0691766) (10.7,0.0697576) (10.8,0.0703397) (10.9,0.0709227) (11,0.0715068) (11.1,0.0720919) (11.2,0.0726779) (11.3,0.0732649) (11.4,0.0738529) (11.5,0.0744417) (11.6,0.0750315) (11.7,0.0756221) (11.8,0.0762136) (11.9,0.0768059) (12,0.0773991) (12.1,0.077993) (12.2,0.0785878) (12.3,0.0791833) (12.4,0.0797797) (12.5,0.0803767) (12.6,0.0809745) (12.7,0.081573) (12.8,0.0821722) (12.9,0.0827721) (13,0.0833727) (13.1,0.0839739) (13.2,0.0845758) (13.3,0.0851783) (13.4,0.0857814) (13.5,0.0863852) (13.6,0.0869895) (13.7,0.0875944) (13.8,0.0881999) (13.9,0.0888059) (14,0.0894125) (14.1,0.0900197) (14.2,0.0906273) (14.3,0.0912355) (14.4,0.0918441) (14.5,0.0924533) (14.6,0.0930629) (14.7,0.093673) (14.8,0.0942836) (14.9,0.0948946) (15,0.0955061) (15.1,0.096118) (15.2,0.0967303) (15.3,0.097343) (15.4,0.0979561) (15.5,0.0985697) (15.6,0.0991836) (15.7,0.0997979) (15.8,0.100413) (15.9,0.101028) (16,0.101643) (16.1,0.102259) (16.2,0.102875) (16.3,0.103491) (16.4,0.104108) (16.5,0.104725) (16.6,0.105342) (16.7,0.10596) (16.8,0.106578) (16.9,0.107196) (17,0.107815) (17.1,0.108433) (17.2,0.109053) (17.3,0.109672) (17.4,0.110292) (17.5,0.110911) (17.6,0.111532) (17.7,0.112152) (17.8,0.112772) (17.9,0.113393) (18,0.114014) (18.1,0.114636) (18.2,0.115257) (18.3,0.115879) (18.4,0.116501) (18.5,0.117123) (18.6,0.117745) (18.7,0.118368) (18.8,0.118991) (18.9,0.119613) (19,0.120236) (19.1,0.12086) (19.2,0.121483) (19.3,0.122107) (19.4,0.122731) (19.5,0.123355) (19.6,0.123979) (19.7,0.124603) (19.8,0.125227) (19.9,0.125852) (20,0.126477)
			};
		\end{axis}
	\end{tikzpicture}
	\caption{Achievable computation rates in a cluster with $N=5$ nodes for the nomographic functions of Examples \ref{ex:arithmetic_mean}--\ref{ex:Euclidean_norm}, where the aimed computation accuracy is set to $\varepsilon=10^{-3}$. The dashed upper bounds correspond with (\ref{eq:AWGN_capacity}).}
	\label{fig:achievable_rates_examples}
\end{figure}
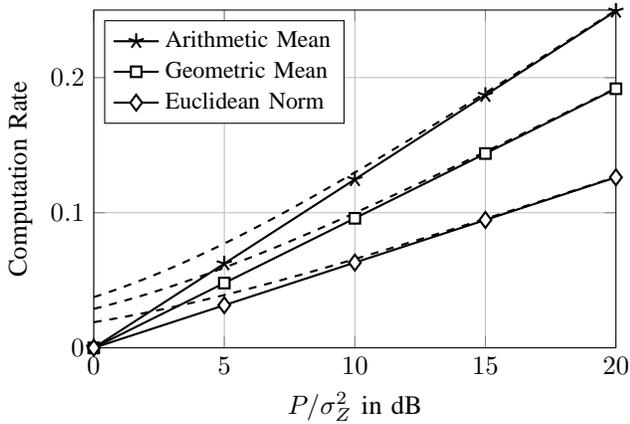%
Fig.~\ref{fig:achievable_rates_examples} depicts the achievable computation rates of Examples \ref{ex:arithmetic_mean}--\ref{ex:Euclidean_norm} for $N=5$, $s_{\text{min}}=10^{-20}$, and $\varepsilon=10^{-3}$ . It turns out, for instance, that at a ratio $P/\sigma_Z^2$ of $15\,\text{dB}$, the ``arithmetic mean'' can be computed approximately $1.3$ times faster than the ``geometric mean'' and approximately $2$ times faster than the ``Euclidean norm'', respectively.

Consider now the standard separation-based computation approach in which the FC reliably decodes all quantized sensor readings individually from the Gaussian MAC output in order to compute the desired function-values afterwards. Then, the corresponding rate performance is limited by the multiple-access capacity region \cite[p.\,98]{ElGamal:Kim:11} from which we conclude that the best computation rate is
\begin{equation}
	R^{\text{Comp}}(f,\varepsilon)=\frac{\frac{1}{2N}\log_2\left(1+\frac{NP}{\sigma_Z^2}\right)}{b_0(f,\varepsilon)}\;,
	\label{eq:multiaccess_rate}
\end{equation}%
which is achievable with Gaussian codebooks in combination with successive cancellation decoding. Comparing (\ref{eq:multiaccess_rate}) with (\ref{eq:max_rate}) reveals that with the scheme of Section~\ref{sec:reliable_computation}, many linear and nonlinear functions can be reliably computed at a rate that is significantly higher than every rate achievable with separation, except for small ratios $P/\sigma_Z^2$. See Section~\ref{sec:discussion} for a more detailed discussion. 

It is easy to see that for $P/\sigma_Z^2\to\infty$, (\ref{eq:max_rate}) achieves an upper bound given by the normalized single-user AWGN capacity:
\begin{equation}
	\bar{R}^{\text{Comp}}(f,\varepsilon)\coloneqq\frac{\frac{1}{2}\log_2\left(1+\frac{P}{\sigma_Z^2}\right)}{b_0(f,\varepsilon)+\log_2(N)}\;.
	\label{eq:AWGN_capacity}
\end{equation}%
Up to date, however, it is unknown whether this bound can also be achieved for finite $P/\sigma_Z^2$. See Fig.~\ref{fig:achievable_rates} for an example.
\begin{figure}[!t]
	\centering
	\begin{tikzpicture}
		\pgfplotsset{every axis legend/.append style={at={(0.02,0.97)},font=\small, anchor=north west, cells={anchor=west}}}
		\pgfplotsset{every axis plot/.append style={thick}}
		\pgfsetplotmarkrepeat{50}
		\begin{axis}[
			width=0.47\textwidth,
			height=0.335\textwidth,
			xmin=0, xmax=20,
			ymin=0, ymax=0.25,
			yticklabel style={/pgf/number format/fixed},
			enlargelimits=false,
			xlabel={$P/\sigma_Z^2$ in dB},
			ylabel={Computation Rate},
			xmajorgrids,
			ymajorgrids,
			legend entries={Computation over MAC (\ref{eq:max_rate}), Successive Decoding (\ref{eq:multiaccess_rate}), Time-Divison Multiple Access}
			]

			\addplot [
			color=black,
			dashed
			]
			plot[forget plot]
			coordinates{
			 (0,0.0349115) (0.1,0.0354947) (0.2,0.0360846) (0.3,0.0366811) (0.4,0.0372844) (0.5,0.0378942) (0.6,0.0385108) (0.7,0.039134) (0.8,0.0397638) (0.9,0.0404002) (1,0.0410432) (1.1,0.0416929) (1.2,0.0423491) (1.3,0.0430118) (1.4,0.0436811) (1.5,0.0443569) (1.6,0.0450391) (1.7,0.0457278) (1.8,0.046423) (1.9,0.0471245) (2,0.0478324) (2.1,0.0485466) (2.2,0.0492672) (2.3,0.049994) (2.4,0.050727) (2.5,0.0514662) (2.6,0.0522116) (2.7,0.0529631) (2.8,0.0537207) (2.9,0.0544842) (3,0.0552538) (3.1,0.0560293) (3.2,0.0568107) (3.3,0.057598) (3.4,0.058391) (3.5,0.0591898) (3.6,0.0599943) (3.7,0.0608044) (3.8,0.0616202) (3.9,0.0624415) (4,0.0632682) (4.1,0.0641004) (4.2,0.064938) (4.3,0.065781) (4.4,0.0666292) (4.5,0.0674826) (4.6,0.0683412) (4.7,0.0692049) (4.8,0.0700736) (4.9,0.0709473) (5,0.071826) (5.1,0.0727095) (5.2,0.0735979) (5.3,0.074491) (5.4,0.0753888) (5.5,0.0762912) (5.6,0.0771982) (5.7,0.0781098) (5.8,0.0790258) (5.9,0.0799462) (6,0.080871) (6.1,0.0818) (6.2,0.0827333) (6.3,0.0836707) (6.4,0.0846122) (6.5,0.0855578) (6.6,0.0865074) (6.7,0.0874609) (6.8,0.0884183) (6.9,0.0893795) (7,0.0903445) (7.1,0.0913132) (7.2,0.0922855) (7.3,0.0932614) (7.4,0.0942408) (7.5,0.0952237) (7.6,0.0962101) (7.7,0.0971998) (7.8,0.0981928) (7.9,0.0991891) (8,0.100189) (8.1,0.101191) (8.2,0.102197) (8.3,0.103206) (8.4,0.104218) (8.5,0.105232) (8.6,0.10625) (8.7,0.10727) (8.8,0.108294) (8.9,0.10932) (9,0.110348) (9.1,0.11138) (9.2,0.112414) (9.3,0.11345) (9.4,0.114489) (9.5,0.115531) (9.6,0.116575) (9.7,0.117621) (9.8,0.11867) (9.9,0.119721) (10,0.120774) (10.1,0.121829) (10.2,0.122887) (10.3,0.123947) (10.4,0.125008) (10.5,0.126072) (10.6,0.127138) (10.7,0.128206) (10.8,0.129276) (10.9,0.130347) (11,0.131421) (11.1,0.132496) (11.2,0.133573) (11.3,0.134652) (11.4,0.135732) (11.5,0.136815) (11.6,0.137899) (11.7,0.138984) (11.8,0.140071) (11.9,0.14116) (12,0.14225) (12.1,0.143342) (12.2,0.144435) (12.3,0.145529) (12.4,0.146625) (12.5,0.147722) (12.6,0.148821) (12.7,0.149921) (12.8,0.151022) (12.9,0.152125) (13,0.153229) (13.1,0.154334) (13.2,0.15544) (13.3,0.156547) (13.4,0.157656) (13.5,0.158765) (13.6,0.159876) (13.7,0.160988) (13.8,0.1621) (13.9,0.163214) (14,0.164329) (14.1,0.165445) (14.2,0.166562) (14.3,0.167679) (14.4,0.168798) (14.5,0.169918) (14.6,0.171038) (14.7,0.172159) (14.8,0.173282) (14.9,0.174405) (15,0.175528) (15.1,0.176653) (15.2,0.177778) (15.3,0.178904) (15.4,0.180031) (15.5,0.181159) (15.6,0.182287) (15.7,0.183416) (15.8,0.184546) (15.9,0.185676) (16,0.186807) (16.1,0.187939) (16.2,0.189071) (16.3,0.190204) (16.4,0.191337) (16.5,0.192471) (16.6,0.193606) (16.7,0.194741) (16.8,0.195877) (16.9,0.197013) (17,0.19815) (17.1,0.199287) (17.2,0.200425) (17.3,0.201563) (17.4,0.202702) (17.5,0.203841) (17.6,0.204981) (17.7,0.206121) (17.8,0.207262) (17.9,0.208403) (18,0.209544) (18.1,0.210686) (18.2,0.211828) (18.3,0.212971) (18.4,0.214114) (18.5,0.215257) (18.6,0.216401) (18.7,0.217545) (18.8,0.21869) (18.9,0.219835) (19,0.22098) (19.1,0.222125) (19.2,0.223271) (19.3,0.224417) (19.4,0.225564) (19.5,0.22671) (19.6,0.227857) (19.7,0.229005) (19.8,0.230152) (19.9,0.2313) (20,0.232448)
			};

			\addplot [
			color=black,
			solid
			]coordinates{
			 (0,0) (0.1,0.00115973) (0.2,0.00231947) (0.3,0.0034792) (0.4,0.00463894) (0.5,0.00579867) (0.6,0.00695841) (0.7,0.00811814) (0.8,0.00927788) (0.9,0.0104376) (1,0.0115973) (1.1,0.0127571) (1.2,0.0139168) (1.3,0.0150766) (1.4,0.0162363) (1.5,0.017396) (1.6,0.0185558) (1.7,0.0197155) (1.8,0.0208752) (1.9,0.022035) (2,0.0231947) (2.1,0.0243544) (2.2,0.0255142) (2.3,0.0266739) (2.4,0.0278336) (2.5,0.0289934) (2.6,0.0301531) (2.7,0.0313128) (2.8,0.0324726) (2.9,0.0336323) (3,0.034792) (3.1,0.0359518) (3.2,0.0371115) (3.3,0.0382713) (3.4,0.039431) (3.5,0.0405907) (3.6,0.0417505) (3.7,0.0429102) (3.8,0.0440699) (3.9,0.0452297) (4,0.0463894) (4.1,0.0475491) (4.2,0.0487089) (4.3,0.0498686) (4.4,0.0510283) (4.5,0.0521881) (4.6,0.0533478) (4.7,0.0545075) (4.8,0.0556673) (4.9,0.056827) (5,0.0579867) (5.1,0.0591465) (5.2,0.0603062) (5.3,0.061466) (5.4,0.0626257) (5.5,0.0637854) (5.6,0.0649452) (5.7,0.0661049) (5.8,0.0672646) (5.9,0.0684244) (6,0.0695841) (6.1,0.0707438) (6.2,0.0719036) (6.3,0.0730633) (6.4,0.074223) (6.5,0.0753828) (6.6,0.0765425) (6.7,0.0777022) (6.8,0.078862) (6.9,0.0800217) (7,0.0811814) (7.1,0.0823412) (7.2,0.0835009) (7.3,0.0846607) (7.4,0.0858204) (7.5,0.0869801) (7.6,0.0881399) (7.7,0.0892996) (7.8,0.0904593) (7.9,0.0916191) (8,0.0927788) (8.1,0.0939385) (8.2,0.0950983) (8.3,0.096258) (8.4,0.0974177) (8.5,0.0985775) (8.6,0.0997372) (8.7,0.100897) (8.8,0.102057) (8.9,0.103216) (9,0.104376) (9.1,0.105536) (9.2,0.106696) (9.3,0.107855) (9.4,0.109015) (9.5,0.110175) (9.6,0.111335) (9.7,0.112494) (9.8,0.113654) (9.9,0.114814) (10,0.115973) (10.1,0.117133) (10.2,0.118293) (10.3,0.119453) (10.4,0.120612) (10.5,0.121772) (10.6,0.122932) (10.7,0.124092) (10.8,0.125251) (10.9,0.126411) (11,0.127571) (11.1,0.128731) (11.2,0.12989) (11.3,0.13105) (11.4,0.13221) (11.5,0.13337) (11.6,0.134529) (11.7,0.135689) (11.8,0.136849) (11.9,0.138008) (12,0.139168) (12.1,0.140328) (12.2,0.141488) (12.3,0.142647) (12.4,0.143807) (12.5,0.144967) (12.6,0.146127) (12.7,0.147286) (12.8,0.148446) (12.9,0.149606) (13,0.150766) (13.1,0.151925) (13.2,0.153085) (13.3,0.154245) (13.4,0.155404) (13.5,0.156564) (13.6,0.157724) (13.7,0.158884) (13.8,0.160043) (13.9,0.161203) (14,0.162363) (14.1,0.163523) (14.2,0.164682) (14.3,0.165842) (14.4,0.167002) (14.5,0.168162) (14.6,0.169321) (14.7,0.170481) (14.8,0.171641) (14.9,0.172801) (15,0.17396) (15.1,0.17512) (15.2,0.17628) (15.3,0.177439) (15.4,0.178599) (15.5,0.179759) (15.6,0.180919) (15.7,0.182078) (15.8,0.183238) (15.9,0.184398) (16,0.185558) (16.1,0.186717) (16.2,0.187877) (16.3,0.189037) (16.4,0.190197) (16.5,0.191356) (16.6,0.192516) (16.7,0.193676) (16.8,0.194835) (16.9,0.195995) (17,0.197155) (17.1,0.198315) (17.2,0.199474) (17.3,0.200634) (17.4,0.201794) (17.5,0.202954) (17.6,0.204113) (17.7,0.205273) (17.8,0.206433) (17.9,0.207593) (18,0.208752) (18.1,0.209912) (18.2,0.211072) (18.3,0.212231) (18.4,0.213391) (18.5,0.214551) (18.6,0.215711) (18.7,0.21687) (18.8,0.21803) (18.9,0.21919) (19,0.22035) (19.1,0.221509) (19.2,0.222669) (19.3,0.223829) (19.4,0.224989) (19.5,0.226148) (19.6,0.227308) (19.7,0.228468) (19.8,0.229628) (19.9,0.230787) (20,0.231947)
			};

			\addplot [
			color=black,
			solid,
			mark=star,
			mark options={scale=1.5,solid,fill=white,thick}
			]coordinates{
			 (0,0.0157247) (0.1,0.0158621) (0.2,0.0159998) (0.3,0.0161378) (0.4,0.016276) (0.5,0.0164145) (0.6,0.0165533) (0.7,0.0166923) (0.8,0.0168316) (0.9,0.0169711) (1,0.0171109) (1.1,0.0172509) (1.2,0.0173911) (1.3,0.0175316) (1.4,0.0176723) (1.5,0.0178132) (1.6,0.0179543) (1.7,0.0180956) (1.8,0.0182372) (1.9,0.0183789) (2,0.0185208) (2.1,0.018663) (2.2,0.0188053) (2.3,0.0189478) (2.4,0.0190905) (2.5,0.0192334) (2.6,0.0193764) (2.7,0.0195196) (2.8,0.019663) (2.9,0.0198066) (3,0.0199503) (3.1,0.0200941) (3.2,0.0202382) (3.3,0.0203823) (3.4,0.0205267) (3.5,0.0206711) (3.6,0.0208157) (3.7,0.0209605) (3.8,0.0211054) (3.9,0.0212504) (4,0.0213955) (4.1,0.0215408) (4.2,0.0216862) (4.3,0.0218318) (4.4,0.0219774) (4.5,0.0221232) (4.6,0.022269) (4.7,0.022415) (4.8,0.0225611) (4.9,0.0227074) (5,0.0228537) (5.1,0.0230001) (5.2,0.0231466) (5.3,0.0232932) (5.4,0.0234399) (5.5,0.0235868) (5.6,0.0237337) (5.7,0.0238807) (5.8,0.0240277) (5.9,0.0241749) (6,0.0243222) (6.1,0.0244695) (6.2,0.0246169) (6.3,0.0247644) (6.4,0.024912) (6.5,0.0250597) (6.6,0.0252074) (6.7,0.0253552) (6.8,0.025503) (6.9,0.025651) (7,0.025799) (7.1,0.0259471) (7.2,0.0260952) (7.3,0.0262434) (7.4,0.0263917) (7.5,0.02654) (7.6,0.0266884) (7.7,0.0268369) (7.8,0.0269854) (7.9,0.0271339) (8,0.0272825) (8.1,0.0274312) (8.2,0.0275799) (8.3,0.0277287) (8.4,0.0278775) (8.5,0.0280264) (8.6,0.0281753) (8.7,0.0283243) (8.8,0.0284733) (8.9,0.0286223) (9,0.0287714) (9.1,0.0289206) (9.2,0.0290697) (9.3,0.029219) (9.4,0.0293682) (9.5,0.0295175) (9.6,0.0296669) (9.7,0.0298162) (9.8,0.0299657) (9.9,0.0301151) (10,0.0302646) (10.1,0.0304141) (10.2,0.0305637) (10.3,0.0307133) (10.4,0.0308629) (10.5,0.0310125) (10.6,0.0311622) (10.7,0.0313119) (10.8,0.0314616) (10.9,0.0316114) (11,0.0317612) (11.1,0.031911) (11.2,0.0320609) (11.3,0.0322107) (11.4,0.0323606) (11.5,0.0325106) (11.6,0.0326605) (11.7,0.0328105) (11.8,0.0329605) (11.9,0.0331105) (12,0.0332605) (12.1,0.0334106) (12.2,0.0335607) (12.3,0.0337108) (12.4,0.0338609) (12.5,0.034011) (12.6,0.0341612) (12.7,0.0343114) (12.8,0.0344616) (12.9,0.0346118) (13,0.034762) (13.1,0.0349123) (13.2,0.0350626) (13.3,0.0352128) (13.4,0.0353631) (13.5,0.0355135) (13.6,0.0356638) (13.7,0.0358141) (13.8,0.0359645) (13.9,0.0361149) (14,0.0362653) (14.1,0.0364157) (14.2,0.0365661) (14.3,0.0367165) (14.4,0.036867) (14.5,0.0370174) (14.6,0.0371679) (14.7,0.0373184) (14.8,0.0374689) (14.9,0.0376194) (15,0.0377699) (15.1,0.0379204) (15.2,0.038071) (15.3,0.0382215) (15.4,0.0383721) (15.5,0.0385226) (15.6,0.0386732) (15.7,0.0388238) (15.8,0.0389744) (15.9,0.039125) (16,0.0392756) (16.1,0.0394262) (16.2,0.0395769) (16.3,0.0397275) (16.4,0.0398781) (16.5,0.0400288) (16.6,0.0401795) (16.7,0.0403301) (16.8,0.0404808) (16.9,0.0406315) (17,0.0407822) (17.1,0.0409329) (17.2,0.0410836) (17.3,0.0412343) (17.4,0.041385) (17.5,0.0415358) (17.6,0.0416865) (17.7,0.0418372) (17.8,0.041988) (17.9,0.0421387) (18,0.0422895) (18.1,0.0424402) (18.2,0.042591) (18.3,0.0427418) (18.4,0.0428925) (18.5,0.0430433) (18.6,0.0431941) (18.7,0.0433449) (18.8,0.0434957) (18.9,0.0436465) (19,0.0437973) (19.1,0.0439481) (19.2,0.0440989) (19.3,0.0442497) (19.4,0.0444006) (19.5,0.0445514) (19.6,0.0447022) (19.7,0.0448531) (19.8,0.0450039) (19.9,0.0451547) (20,0.0453056)
			};

			\addplot [
			color=black,
			solid,
			mark=square*,
			mark options={solid,fill=white,thick}
			]coordinates{
			 (0,0.00454545) (0.1,0.00462139) (0.2,0.00469819) (0.3,0.00477586) (0.4,0.0048544) (0.5,0.00493381) (0.6,0.00501408) (0.7,0.00509522) (0.8,0.00517722) (0.9,0.00526008) (1,0.0053438) (1.1,0.00542838) (1.2,0.00551382) (1.3,0.00560011) (1.4,0.00568725) (1.5,0.00577523) (1.6,0.00586406) (1.7,0.00595373) (1.8,0.00604424) (1.9,0.00613558) (2,0.00622775) (2.1,0.00632074) (2.2,0.00641456) (2.3,0.00650918) (2.4,0.00660462) (2.5,0.00670087) (2.6,0.00679792) (2.7,0.00689576) (2.8,0.0069944) (2.9,0.00709381) (3,0.00719401) (3.1,0.00729498) (3.2,0.00739672) (3.3,0.00749922) (3.4,0.00760247) (3.5,0.00770647) (3.6,0.00781122) (3.7,0.0079167) (3.8,0.00802291) (3.9,0.00812984) (4,0.00823748) (4.1,0.00834584) (4.2,0.00845489) (4.3,0.00856464) (4.4,0.00867508) (4.5,0.00878619) (4.6,0.00889798) (4.7,0.00901043) (4.8,0.00912354) (4.9,0.0092373) (5,0.0093517) (5.1,0.00946673) (5.2,0.00958239) (5.3,0.00969868) (5.4,0.00981557) (5.5,0.00993307) (5.6,0.0100512) (5.7,0.0101698) (5.8,0.0102891) (5.9,0.0104089) (6,0.0105293) (6.1,0.0106503) (6.2,0.0107718) (6.3,0.0108939) (6.4,0.0110165) (6.5,0.0111396) (6.6,0.0112632) (6.7,0.0113874) (6.8,0.011512) (6.9,0.0116372) (7,0.0117628) (7.1,0.0118889) (7.2,0.0120155) (7.3,0.0121426) (7.4,0.0122701) (7.5,0.0123981) (7.6,0.0125265) (7.7,0.0126553) (7.8,0.0127846) (7.9,0.0129144) (8,0.0130445) (8.1,0.013175) (8.2,0.013306) (8.3,0.0134373) (8.4,0.0135691) (8.5,0.0137012) (8.6,0.0138337) (8.7,0.0139665) (8.8,0.0140998) (8.9,0.0142334) (9,0.0143673) (9.1,0.0145016) (9.2,0.0146362) (9.3,0.0147712) (9.4,0.0149064) (9.5,0.015042) (9.6,0.015178) (9.7,0.0153142) (9.8,0.0154507) (9.9,0.0155876) (10,0.0157247) (10.1,0.0158621) (10.2,0.0159998) (10.3,0.0161378) (10.4,0.016276) (10.5,0.0164145) (10.6,0.0165533) (10.7,0.0166923) (10.8,0.0168316) (10.9,0.0169711) (11,0.0171109) (11.1,0.0172509) (11.2,0.0173911) (11.3,0.0175316) (11.4,0.0176723) (11.5,0.0178132) (11.6,0.0179543) (11.7,0.0180956) (11.8,0.0182372) (11.9,0.0183789) (12,0.0185208) (12.1,0.018663) (12.2,0.0188053) (12.3,0.0189478) (12.4,0.0190905) (12.5,0.0192334) (12.6,0.0193764) (12.7,0.0195196) (12.8,0.019663) (12.9,0.0198066) (13,0.0199503) (13.1,0.0200941) (13.2,0.0202382) (13.3,0.0203823) (13.4,0.0205267) (13.5,0.0206711) (13.6,0.0208157) (13.7,0.0209605) (13.8,0.0211054) (13.9,0.0212504) (14,0.0213955) (14.1,0.0215408) (14.2,0.0216862) (14.3,0.0218318) (14.4,0.0219774) (14.5,0.0221232) (14.6,0.022269) (14.7,0.022415) (14.8,0.0225611) (14.9,0.0227074) (15,0.0228537) (15.1,0.0230001) (15.2,0.0231466) (15.3,0.0232932) (15.4,0.0234399) (15.5,0.0235868) (15.6,0.0237337) (15.7,0.0238807) (15.8,0.0240277) (15.9,0.0241749) (16,0.0243222) (16.1,0.0244695) (16.2,0.0246169) (16.3,0.0247644) (16.4,0.024912) (16.5,0.0250597) (16.6,0.0252074) (16.7,0.0253552) (16.8,0.025503) (16.9,0.025651) (17,0.025799) (17.1,0.0259471) (17.2,0.0260952) (17.3,0.0262434) (17.4,0.0263917) (17.5,0.02654) (17.6,0.0266884) (17.7,0.0268369) (17.8,0.0269854) (17.9,0.0271339) (18,0.0272825) (18.1,0.0274312) (18.2,0.0275799) (18.3,0.0277287) (18.4,0.0278775) (18.5,0.0280264) (18.6,0.0281753) (18.7,0.0283243) (18.8,0.0284733) (18.9,0.0286223) (19,0.0287714) (19.1,0.0289206) (19.2,0.0290697) (19.3,0.029219) (19.4,0.0293682) (19.5,0.0295175) (19.6,0.0296669) (19.7,0.0298162) (19.8,0.0299657) (19.9,0.0301151) (20,0.0302646)
			};
		\end{axis}
	\end{tikzpicture}
	\caption{Achievable computation rates in a cluster with $N=10$ nodes if the FC wants to know the ``arithmetic mean'' (see Example \ref{ex:arithmetic_mean}) within accuracy $\varepsilon=10^{-3}$, which requires $b_0(f,\varepsilon)=11\,\text{bit}$. The dashed upper bound represents the single-user AWGN capacity, normalized by $11+\log_2(10)$, whereas time-division multiple access refers to naive time sharing \cite[p. 96]{ElGamal:Kim:11}.}
	\label{fig:achievable_rates}
\end{figure}
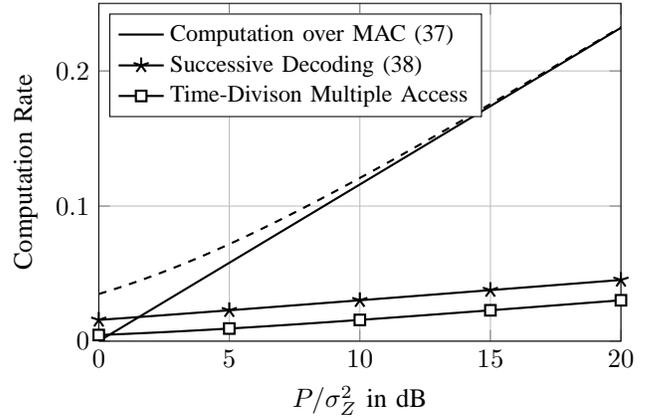%
\begin{remark}\label{rem:log_N}
	The additional logarithmic term in the denominators of (\ref{eq:max_rate}) and (\ref{eq:AWGN_capacity}) is the penalty for avoiding wraparounds in the modulo $p$ sum (\ref{eq:mod_b_messages}) (see Remark~\ref{rem:wraparound}).
\end{remark}%
%
%
%
%
%%%%%%%%%%%%%%%%%%%%%%%%%%%%%%%%%%%%%%%%%%%%%%%%%%%%%%%%%%%%%%%%%%%%%%%%%%%%%%%%%%%%%%%%%%%%%%%%%%%%%%%%%%%%%%%%%%%%%%%%%%%
\subsubsection{Kolmogorov's Superpositions} \label{sec:kolmogorov_single}
\begin{figure*}[!t]
	\centering
	\subfigure[computation transmitter]{
		\begin{picture}(0,0)%
			\includegraphics{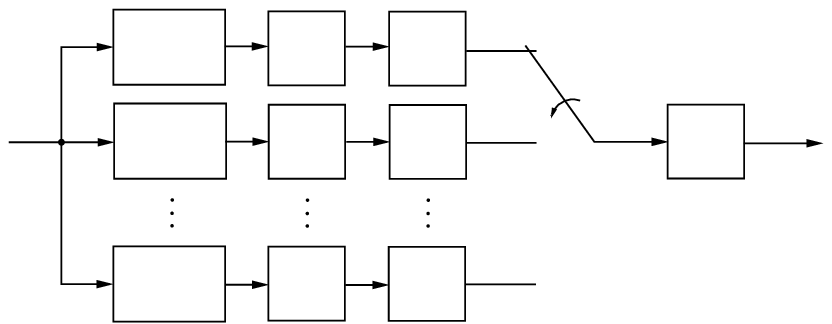}%
		\end{picture}%
		\setlength{\unitlength}{3947sp}%
		\begingroup\makeatletter\ifx\SetFigFont\undefined%
		\gdef\SetFigFont#1#2#3#4#5{%
		\reset@font\fontsize{#1}{#2pt}%
		\fontfamily{#3}\fontseries{#4}\fontshape{#5}%
		\selectfont}%
		\fi\endgroup%
		\begin{picture}(4011,1522)(356,-5415)
			\put(3254,-4486){\makebox(0,0)[lb]{\smash{{\SetFigFont{9}{10.8}{\rmdefault}{\mddefault}{\updefault}{\color[rgb]{0,0,0}$\ve{w}_{ij}$}%
			}}}}
			\put(3728,-4584){\makebox(0,0)[lb]{\smash{{\SetFigFont{9}{10.8}{\rmdefault}{\mddefault}{\updefault}{\color[rgb]{0,0,0}$\mathcal{E}_2$}%
			}}}}
			\put(3999,-4488){\makebox(0,0)[lb]{\smash{{\SetFigFont{9}{10.8}{\rmdefault}{\mddefault}{\updefault}{\color[rgb]{0,0,0}$\ve{x}_{ij}$}%
			}}}}
			\put(1822,-5267){\makebox(0,0)[lb]{\smash{{\SetFigFont{9}{10.8}{\rmdefault}{\mddefault}{\updefault}{\color[rgb]{0,0,0}$Q$}%
			}}}}
			\put(371,-4478){\makebox(0,0)[lb]{\smash{{\SetFigFont{9}{10.8}{\rmdefault}{\mddefault}{\updefault}{\color[rgb]{0,0,0}$s_i[t]$}%
			}}}}
			\put(1112,-4109){\makebox(0,0)[lb]{\smash{{\SetFigFont{9}{10.8}{\rmdefault}{\mddefault}{\updefault}{\color[rgb]{0,0,0}$\varphi_{i1}$}%
			}}}}
			\put(1101,-4567){\makebox(0,0)[lb]{\smash{{\SetFigFont{9}{10.8}{\rmdefault}{\mddefault}{\updefault}{\color[rgb]{0,0,0}$\varphi_{i2}$}%
			}}}}
			\put(969,-5250){\makebox(0,0)[lb]{\smash{{\SetFigFont{9}{10.8}{\rmdefault}{\mddefault}{\updefault}{\color[rgb]{0,0,0}$\varphi_{i,2N+1}$}%
			}}}}
			\put(1821,-4131){\makebox(0,0)[lb]{\smash{{\SetFigFont{9}{10.8}{\rmdefault}{\mddefault}{\updefault}{\color[rgb]{0,0,0}$Q$}%
			}}}}
			\put(1818,-4585){\makebox(0,0)[lb]{\smash{{\SetFigFont{9}{10.8}{\rmdefault}{\mddefault}{\updefault}{\color[rgb]{0,0,0}$Q$}%
			}}}}
			\put(2396,-4140){\makebox(0,0)[lb]{\smash{{\SetFigFont{9}{10.8}{\rmdefault}{\mddefault}{\updefault}{\color[rgb]{0,0,0}$\mathcal{E}_1$}%
			}}}}
			\put(2394,-4594){\makebox(0,0)[lb]{\smash{{\SetFigFont{9}{10.8}{\rmdefault}{\mddefault}{\updefault}{\color[rgb]{0,0,0}$\mathcal{E}_1$}%
			}}}}
			\put(2394,-5274){\makebox(0,0)[lb]{\smash{{\SetFigFont{9}{10.8}{\rmdefault}{\mddefault}{\updefault}{\color[rgb]{0,0,0}$\mathcal{E}_1$}%
			}}}}
			\put(2661,-4057){\makebox(0,0)[lb]{\smash{{\SetFigFont{9}{10.8}{\rmdefault}{\mddefault}{\updefault}{\color[rgb]{0,0,0}$\ve{w}_{i1}$}%
			}}}}
			\put(2666,-4502){\makebox(0,0)[lb]{\smash{{\SetFigFont{9}{10.8}{\rmdefault}{\mddefault}{\updefault}{\color[rgb]{0,0,0}$\ve{w}_{i2}$}%
			}}}}	
			\put(2661,-5175){\makebox(0,0)[lb]{\smash{{\SetFigFont{9}{10.8}{\rmdefault}{\mddefault}{\updefault}{\color[rgb]{0,0,0}$\ve{w}_{i,2N+1}$}%
			}}}}
		\end{picture}%
		\label{fig:computation_transmitter}
		}
	\subfigure[computation receiver]{
		\begin{picture}(0,0)%
			\includegraphics{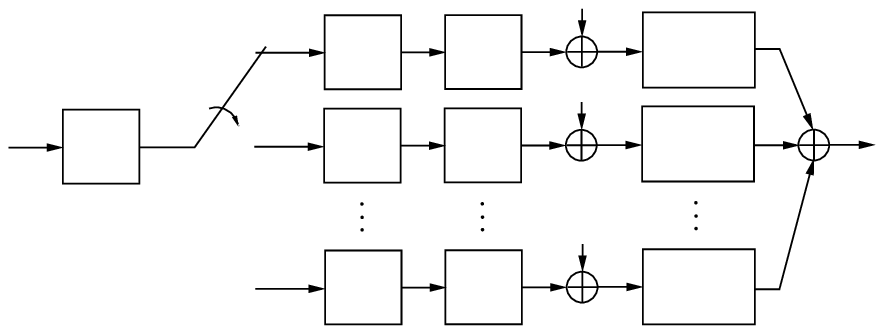}%
		\end{picture}%
		\setlength{\unitlength}{3947sp}%
		\begingroup\makeatletter\ifx\SetFigFont\undefined%
		\gdef\SetFigFont#1#2#3#4#5{%
		\reset@font\fontsize{#1}{#2pt}%
		\fontfamily{#3}\fontseries{#4}\fontshape{#5}%
		\selectfont}%
		\fi\endgroup%
		\begin{picture}(4219,1729)(1011,-5349)
			\put(3207,-4074){\makebox(0,0)[lb]{\smash{{\SetFigFont{9}{10.8}{\rmdefault}{\mddefault}{\updefault}{\color[rgb]{0,0,0}$Q^{-1}$}%
			}}}}
			\put(3206,-4521){\makebox(0,0)[lb]{\smash{{\SetFigFont{9}{10.8}{\rmdefault}{\mddefault}{\updefault}{\color[rgb]{0,0,0}$Q^{-1}$}%
			}}}}
			\put(3210,-5203){\makebox(0,0)[lb]{\smash{{\SetFigFont{9}{10.8}{\rmdefault}{\mddefault}{\updefault}{\color[rgb]{0,0,0}$Q^{-1}$}%
			}}}}
			\put(4984,-4400){\makebox(0,0)[lb]{\smash{{\SetFigFont{9}{10.8}{\rmdefault}{\mddefault}{\updefault}{\color[rgb]{0,0,0}$\hat{f}(\ve{s}_{\ell}[t])$}%
			}}}}
			\put(3701,-3775){\makebox(0,0)[lb]{\smash{{\SetFigFont{9}{10.8}{\rmdefault}{\mddefault}{\updefault}{\color[rgb]{0,0,0}$\gamma_{\ell 1}$}%
			}}}}
			\put(3705,-4231){\makebox(0,0)[lb]{\smash{{\SetFigFont{9}{10.8}{\rmdefault}{\mddefault}{\updefault}{\color[rgb]{0,0,0}$\gamma_{\ell 2}$}%
			}}}}
			\put(3705,-4906){\makebox(0,0)[lb]{\smash{{\SetFigFont{9}{10.8}{\rmdefault}{\mddefault}{\updefault}{\color[rgb]{0,0,0}$\gamma_{\ell,2N+1}$}%
			}}}}
			\put(4262,-4046){\makebox(0,0)[lb]{\smash{{\SetFigFont{9}{10.8}{\rmdefault}{\mddefault}{\updefault}{\color[rgb]{0,0,0}$\psi_{\ell 1}$}%
			}}}}
			\put(4260,-4495){\makebox(0,0)[lb]{\smash{{\SetFigFont{9}{10.8}{\rmdefault}{\mddefault}{\updefault}{\color[rgb]{0,0,0}$\psi_{\ell 2}$}%
			}}}}	
			\put(4115,-5184){\makebox(0,0)[lb]{\smash{{\SetFigFont{9}{10.8}{\rmdefault}{\mddefault}{\updefault}{\color[rgb]{0,0,0}$\psi_{\ell,2N+1}$}%
			}}}}
			\put(1704,-4410){\makebox(0,0)[lb]{\smash{{\SetFigFont{9}{10.8}{\rmdefault}{\mddefault}{\updefault}{\color[rgb]{0,0,0}$\hat{\ve{g}}_{\ell j}$}%
			}}}}
			\put(1415,-4525){\makebox(0,0)[lb]{\smash{{\SetFigFont{9}{10.8}{\rmdefault}{\mddefault}{\updefault}{\color[rgb]{0,0,0}$\mathcal{D}_2$}%
			}}}}
			\put(1026,-4409){\makebox(0,0)[lb]{\smash{{\SetFigFont{9}{10.8}{\rmdefault}{\mddefault}{\updefault}{\color[rgb]{0,0,0}$\ve{y}_{\ell j}$}%
			}}}}
			\put(2294,-3971){\makebox(0,0)[lb]{\smash{{\SetFigFont{9}{10.8}{\rmdefault}{\mddefault}{\updefault}{\color[rgb]{0,0,0}$\hat{\ve{g}}_{\ell 1}$}%
			}}}}
			\put(2282,-4420){\makebox(0,0)[lb]{\smash{{\SetFigFont{9}{10.8}{\rmdefault}{\mddefault}{\updefault}{\color[rgb]{0,0,0}$\hat{\ve{g}}_{\ell 2}$}%
			}}}}
			\put(1988,-5091){\makebox(0,0)[lb]{\smash{{\SetFigFont{9}{10.8}{\rmdefault}{\mddefault}{\updefault}{\color[rgb]{0,0,0}$\hat{\ve{g}}_{\ell,2N+1}$}%
			}}}}
			\put(2671,-4074){\makebox(0,0)[lb]{\smash{{\SetFigFont{9}{10.8}{\rmdefault}{\mddefault}{\updefault}{\color[rgb]{0,0,0}$\mathcal{D}_1$}%
			}}}}
			\put(2677,-4521){\makebox(0,0)[lb]{\smash{{\SetFigFont{9}{10.8}{\rmdefault}{\mddefault}{\updefault}{\color[rgb]{0,0,0}$\mathcal{D}_1$}%
			}}}}
			\put(2679,-5201){\makebox(0,0)[lb]{\smash{{\SetFigFont{9}{10.8}{\rmdefault}{\mddefault}{\updefault}{\color[rgb]{0,0,0}$\mathcal{D}_1$}%
			}}}}
		\end{picture}%
		\label{fig:computation_receiver}
		}
	\caption{Block diagram of the $i$\textsuperscript{th} computation transmitter and the $\ell$\textsuperscript{th} computation receiver (i.e., of FC $\ell$), consisting of adequate data pre- and post-processing as well as of nested lattice encoding and decoding.}
	\label{fig:computation_transceiver}
\end{figure*}

Although Examples \ref{ex:arithmetic_mean}--\ref{ex:Euclidean_norm} demonstrate that $\Nomo^0(\mathds{E}^N)$ contains many functions of practical relevance, it has to be emphasized that by Theorem~\ref{thm:buck2}, $\Nomo^0(\mathds{E}^N)$ is a nowhere dense subset of all continuous functions. By Theorem~\ref{thm:kolmogorov}, however, every real-valued continuous function of $N$ variables can be composed of $2N+1$ elements from $\Nomo^0(\mathds{E}^N)$. We use this to provide in the following the computation rate that is achievable for reliably computing Kolmogorov's superpositions with the scheme depicted in Fig.~\ref{fig:computation_transceiver}. Given some $f\in\C^0(\mathds{E}^N)$, the corresponding estimator is of the form 
\begin{equation}
	\hat{f}\bigl(\ve{s}[t]\bigr) = \sum_{j=1}^{2N+1}\psi_{j}\left(Q^{-1}(\hat{\ve{g}}_{j}[t])\right)\;,\;t=1,\dots,T\;.
	\label{eq:kolmo_estimate}
\end{equation}%
\begin{theorem}\label{thm:kolmogorov_single}
	Given $f\in\C^0(\mathds{E}^N)$, let $\hat{f}$ its estimate defined by (\ref{eq:kolmo_estimate}). Let $\varepsilon>0$ be some given desired accuracy and $b_0(f,\varepsilon)$ be specified as in Lemma~\ref{lem:approximation_error}. Then,
	\begin{equation}
		R^{\textup{Comp}}(f,\varepsilon)=\frac{\frac{1}{4N+2}\log^+_2\left(\frac{P}{\sigma_Z^2}\right)}{b_0(f,\varepsilon)+\log_2(N)}
		\label{eq:rate_kolmogorov}
	\end{equation}%
is an achievable computation rate for $f$ and $\varepsilon$.
\end{theorem}%
\begin{IEEEproof}
	The proof is deferred to Appendix~\ref{app:theorem6}.
\end{IEEEproof}%
\begin{remark}\label{rem:generality}
	Comparing (\ref{eq:rate_kolmogorov}) with (\ref{eq:max_rate}) illustrates that when harnessing the superposition property of the Gaussian MAC, universality with respect to the pre-processing strategy costs additional wireless resources. See Section~\ref{sec:discussion} for a more detailed discussion.
\end{remark}%
%
%
%
%%%%%%%%%%%%%%%%%%%%%%%%%%%%%%%%%%%%%%%%%%%%%%%%%%%%%%%%%%%%%%%%%%%%%%%%%%%%%%%%%%%%%%%%%%%%%%%%%%%%%%%%%%%%%%%%%%%%%%%%%%%
\subsection{The Multiple Cluster Case} \label{sec:multiple_cluster}
Now, consider the general network model introduced in Section~\ref{sec:network}, in which $N$ sensor nodes are divided into $L$ clusters $C_{\ell}$. The objective is to compute at FC $\ell$ some desired function $f_{\ell}\in\C^0(\mathds{E}^{|C_{\ell}|})$ of the associated sensor readings.
%
%%%%%%%%%%%%%%%%%%%%%%%%%%%%%%%%%%%%%%%%%%%%%%%%%%%%%%%%%%%%%%%%%%%%%%%%%%%%%%%%%%%%%%%%%%%%%%%%%%%%%%%%%%%%%%%%%%%%%%%%%%%
\subsubsection{Nomographic Functions} \label{sec:nomographic_multiple}
It can be shown that when restricted to nomographic functions with continuous pre- and post-processing functions, the pre-processing functions can never be chosen to be universal \cite{Goldenbaum:Boche:Stanczak:13b}. This is due to the overlap between clusters. Therefore, the clusters have to be activated in a time-division manner whenever the functions to be computed at adjacent FCs are different. This is illustrated by the corresponding nomographic representations
\begin{equation}
	f_{\ell}\bigl(\ve{s}_{\ell}[t]\bigr)=\psi_{\ell}\left(\sum_{i\in C_{\ell}}\varphi_{i}^{(\ell)}(s_i[t])\right)\;,
	\label{eq:nomo_multi_representation}
\end{equation}%
in which the pre-processing functions depend on $\ell$, $\ell=1,\dots,L$ (i.e., on the FC that is currently addressed). As a consequence, the average computation rate achievable in cluster $\ell$ under a naive time-division strategy scheduling clusters in time follows from Lemma~\ref{lem:approximation_error} and Theorem~\ref{thm:sum_protection} to\footnote{It is assumed that the time is divided into $L$ slots of equal duration.}
\begin{equation}
	\begin{split}
		R^{\text{Comp}}_{\ell}(f_1&,\dots,f_L,\varepsilon)=\\
		&\frac{\frac{1}{2L}\log^+_2\left(\frac{P}{\sigma_Z^2}\right)}{b_0(f_1,\dots,f_L,\varepsilon)+\log_2\bigl(\max_{\ell}|C_{\ell}|\bigr)}\;.
	\end{split}
	\label{eq:nomographic_multiple}
\end{equation}%
In contrast to the single cluster case, the rates depend on $f_1,\dots,f_L$ as the error probability in (\ref{eq:computation_rate_error}) extends in the multi-cluster case to
\begin{equation*}
	\Prob\left(\bigcup_{\ell=1}^L\bigcup_{t=1}^T\left\{\sup_{\ve{s}_{\ell}[t]\in\mathds{E}^{|C_{\ell}|}}\left|\hat{f}_{\ell}\bigl(\ve{s}_{\ell}[t]\bigr)-f_{\ell}\bigl(\ve{s}_{\ell}[t]\bigr)\right|>\varepsilon\right\}\right)\;.
\end{equation*}%
This means that the computation accuracy has to be within $\varepsilon$ for all $\ell$ so that $b_0$ depends on $f_1,\dots,f_L$ (see Remark~\ref{rem:N_dependency}).

Following a similar reasoning for a separation-based approach results in the achievable computation rate
\begin{equation}
	R^{\text{Comp}}_{\ell}(f_1,\dots,f_L,\varepsilon)=\frac{\frac{1}{2L|C_{\ell}|}\log_2\left(1+\frac{|C_{\ell}|P}{\sigma_Z^2}\right)}{b_0(f_1,\dots,f_L,\varepsilon)}\;,
	\label{eq:separation_multiple}
\end{equation}%
which is significantly smaller than (\ref{eq:nomographic_multiple}), except for small ratios $P/\sigma_Z^2$.
\begin{remark}\label{rem:rates_equal}
	Observe that (\ref{eq:nomographic_multiple}) is independent of $\ell$ and therefore equal for all clusters.
\end{remark}%
\begin{remark}\label{rem:higher}
	In clustered networks in which $C_{\ell}\cap C_{\ell'}\neq\varnothing$ for all $\ell,\ell'$, (\ref{eq:nomographic_multiple}) and (\ref{eq:separation_multiple}) cannot be increased by more clever time-sharing. The reason is that the common nodes transmit continuously and would therefore violate the average power constraint when increasing their transmit powers by a factor of $L$. If, on the other hand, some of the clusters are disjoint (see Fig.~\ref{fig:clustered_network} for an example), the rates could be improved by designing a time-division protocol that activates these clusters simultaneously. This, however, would further increase the coordination effort and is out of the scope of this paper.
\end{remark}%
%
%
%
%
%%%%%%%%%%%%%%%%%%%%%%%%%%%%%%%%%%%%%%%%%%%%%%%%%%%%%%%%%%%%%%%%%%%%%%%%%%%%%%%%%%%%%%%%%%%%%%%%%%%%%%%%%%%%%%%%%%%%%%%%%%%
\subsubsection{Kolmogorov's Superpositions} \label{sec:kolmogorov_multiple}
Since the pre-processing functions in (\ref{eq:kolmogorov_cluster_alt}) are universal, the function FC $\ell$ computes is determined by the choice of the post-processing functions $\{\psi_{\ell j}\}_{j=1}^{2N+1}$ only. As a consequence, the data pre-processing and lattice encoding is fixed so that an additional protocol for coordinating the activation of clusters, as it was required for achieving (\ref{eq:nomographic_multiple}) and (\ref{eq:separation_multiple}), is not necessary. Thus, the computation rate achievable with the scheme of Fig.~\ref{fig:computation_transceiver}, follows from Theorem~\ref{thm:kolmogorov_single} to
\begin{equation}
	\begin{split}
		R^{\text{Comp}}_{\ell}(f_1&,\dots,f_L,\varepsilon)=\\
		&\frac{\frac{1}{4N+2}\log^+_2\left(\frac{P}{\sigma_Z^2}\right)}{b_0(f_1,\dots,f_L,\varepsilon)+\log_2\bigl(\max_{\ell}|C_{\ell}|\bigr)}\;,
	\end{split}
	\label{eq:kolmogorov_multiple}
\end{equation}%
for all $\ell=1,\dots,L$.

The attentive reader might note that according to Theorem~\ref{thm:kolmogorov}, every continuous desired function $f_{\ell}$ can be represented as
\begin{equation}
	f_{\ell}\bigl(\ve{s}_{\ell}[t]\bigr)=\sum_{j=1}^{2|C_{\ell}|+1}\psi_{\ell j}\left(\sum_{i\in C_{\ell}}\varphi_{ij}^{(\ell)}\bigl(s_i[t]\bigr)\right)\;,
	\label{eq:kolmogorov_cluster}
\end{equation}%
which apparently requires less pre- and post-processing functions than representation (\ref{eq:kolmogorov_cluster_alt}). The reason for preferring (\ref{eq:kolmogorov_cluster_alt}), however, is that due to the coupling between clusters, the pre-processing functions in (\ref{eq:kolmogorov_cluster}) depend on $\ell$. In order to illustrate this please recap from Remark \ref{rem:geometry} that there exists for each $\ell$ a homeomorphism $\bigl(s_{\ell_1},\dots,s_{\ell_{|C_{\ell}|}}\bigr)\mapsto\bigl(\sum_{i\in C_{\ell}}\varphi_{i1}(s_{\ell_i}),\dots,\sum_{i\in C_{\ell}}\varphi_{i,2|C_{\ell}|+1}(s_{\ell_i})\bigr)$ between $\mathds{E}^{|C_{\ell}|}$ and some compact $\Gamma_{\ell}\subset\mathds{R}^{2|C_{\ell}|+1}$, which allows every $f_{\ell}\in\C^0(\mathds{E}^{|C_{\ell}|})$ to be written as in (\ref{eq:kolmogorov_cluster}) with the same universal set of pre-processing functions. Since the $\Gamma_{\ell}$ will differ in general, the pre-processing functions in (\ref{eq:kolmogorov_cluster}) depend on $\ell$. Whereas this is irrelevant for nodes whose transmit signals can be received by only a single FC, the common nodes that can be heard by more than one FC have to adapt their pre-processing in accordance to the FC that they want to address in a subsequent step. Hence, the coordinated activation of clusters in a time-division manner would be necessary as it was already the case for achieving (\ref{eq:nomographic_multiple}) so that the corresponding achievable computation rate in cluster $\ell$, $\ell=1,\dots,L$, would follow from Theorem~\ref{thm:kolmogorov_single} to
\begin{equation}
	\begin{split}
		R^{\text{Comp}}_{\ell}(f_1&,\dots,f_L,\varepsilon)=\\
		&\frac{\frac{1}{(4|C_{\ell}|+2)L}\log^+_2\left(\frac{P}{\sigma_Z^2}\right)}{b_0(f_1,\dots,f_L,\varepsilon)+\log_2\bigl(\max_{\ell}|C_{\ell}|\bigr)}\;.
	\end{split}
	\label{eq:kolmo_rate_suboptimal}
\end{equation}%
A comparison of (\ref{eq:kolmo_rate_suboptimal}) with (\ref{eq:kolmogorov_multiple}) shows that this computation strategy would lead to a superior rate performance only in those clusters in which $(2|C_{\ell}|+1)L<2N+1$ applies, but at the cost of additional coordination.
%
% 
%
%
%%%%%%%%%%%%%%%%%%%%%%%%%%%%%%%%%%%%%%%%%%%%%%%%%%%%%%%%%%%%%%%%%%%%%%%%%%%%%%%%%%%%%%%%%%%%%%%%%%%%%%%%%%%%%%%%%%%%%%%%%%%
\subsection{Discussion of the Results} \label{sec:discussion}
The results for the single cluster case in Section~\ref{sec:single_cluster} show that when harnessing the superposition property of the Gaussian MAC, nomographic functions with continuous pre- and post-processing functions can be computed significantly faster than with any separation-based strategy, which was illustrated in Fig.~\ref{fig:achievable_rates}. When considering the computation of arbitrary continuous functions of the sensor readings, the corresponding computation rates scale down by a factor of $2N+1$ (see (\ref{eq:rate_kolmogorov})).

In a network of multiple clusters, as shown in Section~\ref{sec:multiple_cluster}, the computation rate achievable when considering in each cluster an individual continuous nomographic function is reduced by a factor of $L$ (see (\ref{eq:nomographic_multiple}) and (\ref{eq:separation_multiple})) since additional coordination is necessary in the form of time sharing between clusters. In contrast, due to the universality of pre-processing functions and the particular data post-processing strategy depicted in Fig.~\ref{fig:computation_transceiver}, the rate at which in each cluster an individual Kolmogorov's superposition can be computed is further given by Theorem~\ref{thm:kolmogorov_single}, \emph{regardless} of the coupling between clusters.

In the domain of wireless sensor networks, achieving high rates is generally not the main concern. Due to limited energy and processing capabilities, computation schemes of low complexity are also of particular interest. Considering the results of Section~\ref{sec:multiple_cluster} from this perspective reveals that the proposed computation scheme has not only in the case of continuous nomographic functions several advantages over separation-based approaches. For example, when computing a set of individual Kolmogorov's superpositions in a clustered network, any coordination of nodes or clusters is not necessary as it would be the case for continuous nomographic functions or separation-based approaches. Especially for large networks with many clusters, this may lead to significant savings in complexity so that computing Kolmogorov's superpositions (i.e., continuous functions) over the channel can be an option even if the achievable computation rate is not maximal. It is clear from the structure of nomographic functions, and therefore Kolmogorov's superpositions, that the computation of only one-variable functions is required at FCs, which can be less demanding than computing the multivariate desired function given the entire set of raw sensor readings, such as in the case of separation-based computation.

If the underlying application is satisfied with the computation of continuous nomographic functions, then in addition to the superior rate performance, the scheme proposed in this paper has a significantly lower decoding complexity, which is essentially the complexity of a single-user lattice decoder. As a consequence, the decoding complexity in cluster $\ell$, $\ell=1,\dots,L$, is $|C_{\ell}|$-fold less than for separation-based computation in which the FC has to reliably decode all the sensor readings gathered in cluster $C_{\ell}$. The latter has also the drawback of a higher sensitivity regarding decoding errors since already a single wrongly decoded sensor reading results in a faulty function-value. We would like to emphasize that the computation rates presented in this section are all achievable under a maximum probability of error criterion as it is indispensable for most sensor network applications.

When additionally using common randomness at sensor nodes and FCs in combination with minimum mean square error estimation prior to decoding \cite{Nazer:Gastpar:11b,Wilson:Narayanan:Pfister:Sprintson:10}, slightly higher computation rates could be achieved than those presented in Theorems~\ref{thm:sum_protection} and \ref{thm:kolmogorov_single}. This, however, would have the drawback that only average error probabilities could be handled with uniformly distributed sensor readings.
%
%
%
%
%%%%%%%%%%%%%%%%%%%%%%%%%%%%%%%%%%%%%%%%%%%%%%%%%%%%%%%%%%%%%%%%%%%%%%%%%%%%%%%%%%%%%%%%%%%%%%%%%%%%%%%%%%%%%%%%%%%%%%%%%%%
%%%%%%%%%%%%%%%%%%%%%%%%%%%%%%%%%%%%%%%%%%%%%%%%%%%%%%%%%%%%%%%%%%%%%%%%%%%%%%%%%%%%%%%%%%%%%%%%%%%%%%%%%%%%%%%%%%%%%%%%%%%
%	Conclusion
%%%%%%%%%%%%%%%%%%%%%%%%%%%%%%%%%%%%%%%%%%%%%%%%%%%%%%%%%%%%%%%%%%%%%%%%%%%%%%%%%%%%%%%%%%%%%%%%%%%%%%%%%%%%%%%%%%%%%%%%%%%
%%%%%%%%%%%%%%%%%%%%%%%%%%%%%%%%%%%%%%%%%%%%%%%%%%%%%%%%%%%%%%%%%%%%%%%%%%%%%%%%%%%%%%%%%%%%%%%%%%%%%%%%%%%%%%%%%%%%%%%%%%%
%
\section{Conclusion} \label{sec:conclusion}
In this work, we considered the reliable computation of arbitrary continuous functions of the measurements in clustered Gaussian sensor networks. In doing so, it has been found that when appropriately harnessing the superposition property of the underlying Gaussian multiple-access channels, a certain subset of all continuous functions (i.e., the set of continuous nomographic functions) can be computed at considerably higher rates than those achievable with an approach that intends to decode all associated sensor readings at the fusion centers for computing the function-values afterwards. Since many continuous functions of practical relevance are nomographic, the result extends the known results for the computation of linear functions to numerous nonlinear functions. 

When the computation of \emph{arbitrary} continuous functions is desired, then the presented approach that combines a suitable data pre- and post-processing strategy with a simple quantizer and nested lattice codes requires the successive computation of multiple nomographic functions, which scales down the achievable computation rates accordingly. Even though these rates can be inferior to those achievable with standard multiple-access schemes, the proposed approach provides several other advantages that are indispensable in many sensor network applications such as lower decoding complexity and less coordination burden. As a consequence, the results of this paper partially carry over the results of \cite{Goldenbaum:Boche:Stanczak:13b} to noisy networks. 

Note that the clustered Gaussian sensor network model considered in this paper assumes that the channel gains between nodes and FCs are all equal to one (see (\ref{eq:WMAC})). For many applications, however, the propagation conditions are more challenging as corresponding wireless transmissions may be subject to fading effects. In networks with non-overlapping clusters this is not a big issue as nodes could invert their channels by employing channel state information. In contrast, when clusters allowed to overlap, some of the nodes can be heard by more than one FC, which generally results in different channel gains. Since this can have a detrimental impact on the computation rate performance, it has to be figured out in future work how to appropriately cope with this.
%
%
%
%%%%%%%%%%%%%%%%%%%%%%%%%%%%%%%%%%%%%%%%%%%%%%%%%%%%%%%%%%%%%%%%%%%%%%%%%%%%%%%%%%%%%%%%%%%%%%%%%%%%%%%%%%%%%%%%%%%%%%%%%%%
%%%%%%%%%%%%%%%%%%%%%%%%%%%%%%%%%%%%%%%%%%%%%%%%%%%%%%%%%%%%%%%%%%%%%%%%%%%%%%%%%%%%%%%%%%%%%%%%%%%%%%%%%%%%%%%%%%%%%%%%%%%
%	Appendix
%%%%%%%%%%%%%%%%%%%%%%%%%%%%%%%%%%%%%%%%%%%%%%%%%%%%%%%%%%%%%%%%%%%%%%%%%%%%%%%%%%%%%%%%%%%%%%%%%%%%%%%%%%%%%%%%%%%%%%%%%%%
%%%%%%%%%%%%%%%%%%%%%%%%%%%%%%%%%%%%%%%%%%%%%%%%%%%%%%%%%%%%%%%%%%%%%%%%%%%%%%%%%%%%%%%%%%%%%%%%%%%%%%%%%%%%%%%%%%%%%%%%%%%
%
\appendix
%
%%%%%%%%%%%%%%%%%%%%%%%%%%%%%%%%%%%%%%%%%%%%%%%%%%%%%%%%%%%%%%%%%%%%%%%%%%%%%%%%%%%%%%%%%%%%%%%%%%%%%%%%%%%%%%%%%%%%%%%%%%%
\subsection{Proof of Lemma~\ref{lem:approximation_error}} \label{app:lemma3}
Observe that an expansion in the way of (\ref{eq:approx_phi}) represents along with $v=\lfloor\log_2(\pi_{\text{max}})\rfloor$ the pre-processed sensor readings up to precision
\begin{equation}
	\bigl|\varphi_{ij}(s)-\tilde{\varphi}_{ij}(s)\bigr|<2^{-\eta}=2^{-b+v+1}\leq\pi_{\text{max}}2^{-b+1}\;,
	\label{eq:phi_approx}
\end{equation}%
for all $s\in\mathds{E}$, $i=1,\dots,N$, and $j=1,\dots,2N+1$. Hence, we can bound the accuracy of the sum of pre-processed sensor readings by virtue of the triangle inequality to
\begin{equation}
	\begin{split}
		\biggl|\sum_{i\in C_{\ell}}&\varphi_{ij}(s_i)-\sum_{i\in C_{\ell}}\tilde{\varphi}_{ij}(s_i)\biggr|\\
		&\leq
	\sum_{i\in C_{\ell}}\bigl|\varphi_{ij}(s_i)-\tilde{\varphi}_{ij}(s_i)\bigr|<|C_{\ell}|\pi_{\text{max}}2^{-b+1}\;,
	\end{split}
	\label{eq:sum_phi_approx}
\end{equation}%
for all $\ve{s}_{\ell}\in\mathds{E}^{|C_{\ell}|}$, $\ell=1,\dots,L$, and $j=1,\dots,2N+1$. 

Since the constants in (\ref{eq:constants}) are bounded and continuity on compact metric spaces implies uniform continuity \cite[p.\,91]{Rudin:76}, we conclude from (\ref{eq:sum_phi_approx}) that\footnote{Because every finite sum of compact spaces is compact, it follows from the compactness of the $\Pi_{ij}$ (i.e., the ranges of pre-processing functions) that the ranges of the sums $\sum_{i\in C_{\ell}}\varphi_{ij}(s_i)$, $\ell=1,\dots,L$; $j=1,\dots,2N+1$, are compact as well.}
\begin{equation}
	\begin{split}
		\sup_{\ve{s}_{\ell}\in\mathds{E}^{|C_{\ell}|}}\Biggl|\psi_{\ell j}&\left(\sum_{i\in C_{\ell}}\varphi_{ij}(s_i)+\gamma_{\ell j}\right)\\
		&-\psi_{\ell j}\left(\sum_{i\in C_{\ell}}\tilde{\varphi}_{ij}(s_i)+\gamma_{\ell j}\right)\Biggr|<\varepsilon_{\ell j}(b)\;,
	\end{split}
	\label{eq:approx_psi}
\end{equation}%
for some $\varepsilon_{\ell j}(b)>0$ and for all $\ell,j$. Now, let $\varepsilon>0$ be arbitrary but fixed. Then, there exists $b_0=b_0(f_1,\dots,f_L,N,\varepsilon)$ such that $\max_{\ell,j}\varepsilon_{\ell,j}(b)<\frac{\varepsilon}{2N+1}$, for all $b\geq b_0$. As a consequence, we have for all $\ve{s}_{\ell}\in\mathds{E}^{|C_{\ell}|}$, $\ell=1,\dots,L$, and $b\geq b_0$
\begin{equation*}
	\begin{split}
		\left|f_{\ell}(\ve{s}_{\ell})-\tilde{f}_{\ell}(\ve{s}_{\ell})\right|\leq &\sum_{j=1}^{2N+1}\Biggl|\psi_{\ell j}\left(\sum_{i\in C_{\ell}}\varphi_{ij}(s_i)+\gamma_{\ell j}\right)\\
		&\hspace{14pt}-\psi_{\ell j}\left(\sum_{i\in C_{\ell}}\tilde{\varphi}_{ij}(s_i)+\gamma_{\ell j}\right)\Biggr|<\varepsilon\;,
	\end{split}
	\label{eq:overall_approx}
\end{equation*}%
which proves the lemma.
%
%
%
%%%%%%%%%%%%%%%%%%%%%%%%%%%%%%%%%%%%%%%%%%%%%%%%%%%%%%%%%%%%%%%%%%%%%%%%%%%%%%%%%%%%%%%%%%%%%%%%%%%%%%%%%%%%%%%%%%%%%%%%%%%
\subsection{Proof of Theorem~\ref{thm:sum_protection}} \label{app:theorem5}
For some $T\in\mathds{N}$ to be specified below, consider the sequence $\{f(\ve{s}[t])=\psi(\sum_{i=1}^N\varphi_i(s_i[t]))\}_{t=1}^T$ of nomographic function-values with continuous pre- and post-processing functions and let $\{\tilde{f}(\ve{s}[t])\coloneqq\psi(\sum_{i=1}^N\tilde{\varphi}_i(s_i[t]))\}_{t=1}^T$ denote the corresponding approximations in accordance with (\ref{eq:quantized_sum_pre}). In addition, let $\varepsilon>0$ be arbitrary but fixed and choose the quantization parameter $b=b(f,\varepsilon)$ as in Lemma~\ref{lem:approximation_error} to $b_0=b_0(f,\varepsilon)$ so that $\sup_{\ve{s}\in\mathds{E}^N}|f(\ve{s})-\tilde{f}(\ve{s})|<\varepsilon$. 

In order to prove the theorem, we have to first construct the source encoder (\ref{eq:source_encoder}). To this end, each of the binary representations (\ref{eq:messages}) is equivalently considered as an element of the set of integers $\{0,1,\dots,2^{b_0}-1\}$, which we denote in the following as $w_i[t]$ to avoid confusion with the vector notation. With this in mind, for each $t\in\{1,\dots,T\}$, the sum of these integers is bounded above as
\begin{equation}
	\sum_{i=1}^N w_i[t]\leq N(2^{b_0}-1)\eqqcolon q-1\;.
	\label{eq:sum_bound}
\end{equation}%
Now, for some $\tau\in\mathds{N}$ to be specified below, we form the length-$k$ messages (\ref{eq:overall_messages}) in the following way
\begin{equation*}
	\ve{w}_i=\left(\sum_{t=1}^{\tau}w_i[t]q^{t-1},\dots,\sum_{t=1}^{\tau}w_i[t+(k-1)\tau]q^{t-1}\right),
	\label{eq:messages_constructed}
\end{equation*}% 
$i=1,\dots,N$, with $q$ as defined in (\ref{eq:sum_bound}). Note that the sum over $i$ of each component is bounded above as
\begin{equation}
	\sum_{i=1}^N\sum_{t=1}^{\tau}w_i[t]q^{t-1}=\sum_{t=1}^{\tau}\sum_{i=1}^Nw_i[t]q^{t-1}\leq q^{\tau}-1\;.
\end{equation}%
Hence, for every fixed $q$ and alphabet size $p$ (see (\ref{eq:source_encoder}) and (\ref{eq:encoder})), choosing $\tau$ such that
\begin{equation}
	q^{\tau}-1\leq p-1
	\label{eq:tau}
\end{equation}%
avoids any wraparound when messages add up over the channel. Thus, with the right hand side of (\ref{eq:sum_bound}) we have
\begin{equation}
	\tau\leq\frac{\log_2(p)}{\log_2\bigl(2^{b_0(f,\varepsilon)}-1\bigr)+\log_2(N)}\;.
	\label{eq:tau_bound}
\end{equation}%
Now, consider the more conservative bound
\begin{equation}
	\tau\leq\frac{\log_2(p)}{b_0(f,\varepsilon)+\log_2(N)}
	\label{eq:tau_bound2}
\end{equation}%
by ignoring the $-1$ in the denominator of (\ref{eq:tau_bound}). Then, as the number of encoded sensor readings is $T=k\tau$, we conclude for the computation rate (see Definition~\ref{def:computation_rate})
\begin{equation*}
	R'=\frac{T}{n}\leq\frac{k\log_2(p)}{n\bigl(b_0(f,\varepsilon)+\log_2(N)\bigr)}=\frac{R}{b_0(f,\varepsilon)+\log_2(N)}\;.
\end{equation*}%

Note that the computation scheme proposed in Section~\ref{sec:reliable_computation} employs a code-sequence based on a sequence of nested lattices chosen from Lemma~\ref{lem:erez}. As a consequence, $P_e^{(n)}\leq\Prob(\ve{z}\notin\mathcal{V}_{\text{c}}^{(n)})\to 0$ exponentially fast in $n$ as long as the message rate (\ref{eq:message_rate}) fulfills at each node
\setlength{\arraycolsep}{0.0em}
\begin{eqnarray}
	R=\frac{k}{n}\log_2(p)&{}={}&\frac{1}{n}\log_2\left(\frac{\Vol(\mathcal{V}_{\text{s}}^{(n)})}{\Vol(\mathcal{V}_{\text{c}}^{(n)})}\right)\nonumber\\
	&{}\overset{(a)}{=}{}&\frac{1}{2}\log_2\left(\frac{P}{G(\Lambda_{\text{s}}^{(n)})\Vol(\mathcal{V}_{\text{c}}^{(n)})^{2/n}}\right)\nonumber\\
	&{}\overset{(b)}{<}{}&\frac{1}{2}\log_2^+\left(\frac{P}{\sigma_Z^2}\right)-\frac{1}{2}\log_2\bigl(2\pi\!\e G(\Lambda_{\text{s}}^{(n)})\bigr)\;.\nonumber
	\label{eq:comp_rate_bound}
\end{eqnarray}%
\setlength{\arraycolsep}{5pt}%
Here, $(a)$ follows from Definition~\ref{def:second_moment} and the fact that each shaping lattice $\Lambda_{\text{s}}^{(n)}$ is scaled such that its second moment equals the power constraint $P$ whereas $(b)$ is a consequence of the sequence of coding lattices $\{\Lambda_{\text{c}}^{(n)}\}$ being good for AWGN channel coding (see Definition~\ref{def:goodness}). Because the sequence $\{\Lambda_{\text{s}}^{(n)}\}$ is simultaneously good for shaping (i.e., $\lim_{n\to\infty}\log_2(2\pi\!\e G(\Lambda_{\text{s}}^{(n)}))=0$), we therefore have $P_e^{(n)}\to 0$ exponentially fast with growing $n$ if
\begin{equation*}
	R<\frac{1}{2}\log^+_2\left(\frac{P}{\sigma_Z^2}\right)\;.
\end{equation*}%
Consequently, letting $T$, and thus $k$ and $p$, grow appropriately with $n$, expression
\begin{equation}
	\Prob\left(\bigcup_{t=1}^T\Bigl\{\hat{g}(\ve{s}[t])\neq\tilde{g}(\ve{s}[t])\Bigr\}\right)
	\label{eq:error_prob}
\end{equation}%
vanishes exponentially fast in $n$ as well, provided that
\begin{equation}
	R'<\frac{\frac{1}{2}\log_2^+\left(\frac{P}{\sigma_Z^2}\right)}{b_0(f,\varepsilon)+\log_2(N)}=R^{\text{Comp}}(f,\varepsilon)\;,
	\label{eq:comp_rate_nomo}
\end{equation}%
where $\tilde{g}(\ve{s}[t])=\sum_{i=1}^N\tilde{\varphi}_i(s_i[t])$ and $\hat{g}(\ve{s}[t])=Q^{-1}(\hat{\ve{g}}[t])$ are the corresponding estimate at the FC.

Now, recall that $\mathcal{D}_1(\hat{\ve{g}})=(\hat{\ve{g}}[1],\dots,\hat{\ve{g}}[T])$ and choose $\hat{\ve{g}}[t]$ for some fixed $t\in\{1,\dots,T\}$ such that $\psi\bigl(\hat{g}(\ve{s}[t])\bigr)\neq\psi\bigl(\tilde{g}(\ve{s}[t])\bigr)$. Then, this choice implies $\hat{g}(\ve{s}[t])\neq \tilde{g}(\ve{s}[t])$ because $\psi$ is a function. Summarizing all such outage events in the sets $A\coloneqq\left\{\hat{\ve{g}}[t]\,|\,\hat{g}(\ve{s}[t])\neq \tilde{g}(\ve{s}[t])\right\}$ and $B\coloneqq\left\{\hat{\ve{g}}[t]\,|\,\psi\bigl(\hat{g}(\ve{s}[t])\bigr)\neq\psi\bigl(\tilde{g}(\ve{s}[t])\bigr)\right\}$, we have $B\subseteq A$ and therefore $\Prob(B)\leq\Prob(A)$ due to the monotonicity of probability and the measurability of $\psi$. Hence, we can conclude from (\ref{eq:error_prob}) that for each $t\in\{1,\dots,T\}$,
\begin{equation*}
	\Prob\Bigl(\psi\bigl(\hat{g}(\ve{s}[t])\bigr)\neq\psi\bigl(\tilde{g}(\ve{s}[t])\bigr)\Bigr)=\Prob\bigl(\hat{f}(\ve{s}[t])\neq\tilde{f}(\ve{s}[t])\bigr)
\end{equation*}%
goes to zero exponentially fast in $n$, regardless of the choice of $\ve{s}[t]\in\mathds{E}^N$. Since almost sure convergence implies convergence in probability, we therefore have for every $\delta>0$ that $\sum_{t=1}^T\Prob(\sup_{\ve{s}[t]\in\mathds{E}^N}|\hat{f}(\ve{s}[t])-f(\ve{s}[t])|>\varepsilon)<\delta$ if $n$ is sufficiently large, which implies (\ref{eq:computation_rate_error}) due to the union bound.

From this, we conclude that the function-values $f(\ve{s}[t])$ can be computed with high probability within accuracy $\varepsilon$ at a computation rate that is as close to the right hand side of (\ref{eq:comp_rate_nomo}) as desired. This proves the theorem.
%
%
%
%%%%%%%%%%%%%%%%%%%%%%%%%%%%%%%%%%%%%%%%%%%%%%%%%%%%%%%%%%%%%%%%%%%%%%%%%%%%%%%%%%%%%%%%%%%%%%%%%%%%%%%%%%%%%%%%%%%%%%%%%%%
\subsection{Proof of Theorem~\ref{thm:kolmogorov_single}} \label{app:theorem6}
Representing $f$ as its Kolmogorov's superposition (see Theorem~\ref{thm:kolmogorov}) suggests that it can be computed at the FC by successively computing the corresponding $2N+1$ nomographic functions over the Gaussian MAC. Hence, given some fixed $\varepsilon>0$, choose $b=b(f,\varepsilon)$ in accordance with Lemma~\ref{lem:approximation_error} to $b_0$ so that the quantization error is smaller than $\varepsilon$. Now, due to (\ref{eq:error_prob_bound}), we have for the decoding error probability at the FC $P_e^{(n)}\leq\sum_{j=1}^{2N+1}\Prob(\hat{\ve{g}}_{j}\neq\ve{g}_{j})$. Therefore, the theorem follows from Theorem~\ref{thm:sum_protection} by taking into account that for each $j$, $\Prob(\hat{\ve{g}}_j\neq\ve{g}_j)$ goes to zero exponentially fast in the block length $n$ as long as (\ref{eq:max_rate}) is fulfilled.
%
%
%
%%%%%%%%%%%%%%%%%%%%%%%%%%%%%%%%%%%%%%%%%%%%%%%%%%%%%%%%%%%%%%%%%%%%%%%%%%%%%%%%%%%%%%%%%%%%%%%%%%%%%%%%%%%%%%%%%%%%%%%%%%%
%%%%%%%%%%%%%%%%%%%%%%%%%%%%%%%%%%%%%%%%%%%%%%%%%%%%%%%%%%%%%%%%%%%%%%%%%%%%%%%%%%%%%%%%%%%%%%%%%%%%%%%%%%%%%%%%%%%%%%%%%%%
%	Acknowledgments
%%%%%%%%%%%%%%%%%%%%%%%%%%%%%%%%%%%%%%%%%%%%%%%%%%%%%%%%%%%%%%%%%%%%%%%%%%%%%%%%%%%%%%%%%%%%%%%%%%%%%%%%%%%%%%%%%%%%%%%%%%%
%%%%%%%%%%%%%%%%%%%%%%%%%%%%%%%%%%%%%%%%%%%%%%%%%%%%%%%%%%%%%%%%%%%%%%%%%%%%%%%%%%%%%%%%%%%%%%%%%%%%%%%%%%%%%%%%%%%%%%%%%%%
%
\section*{Acknowledgment}
The authors would like to thank the anonymous reviewers for their comments and suggestions, which helped to improve the quality of this paper. The first author would also like to thank Micha{\l} Kaliszan for many helpful discussions.
%
%
%
%%%%%%%%%%%%%%%%%%%%%%%%%%%%%%%%%%%%%%%%%%%%%%%%%%%%%%%%%%%%%%%%%%%%%%%%%%%%%%%%%%%%%%%%%%%%%%%%%%%%%%%%%%%%%%%%%%%%%%%%%%%
%%%%%%%%%%%%%%%%%%%%%%%%%%%%%%%%%%%%%%%%%%%%%%%%%%%%%%%%%%%%%%%%%%%%%%%%%%%%%%%%%%%%%%%%%%%%%%%%%%%%%%%%%%%%%%%%%%%%%%%%%%%
%	References
%%%%%%%%%%%%%%%%%%%%%%%%%%%%%%%%%%%%%%%%%%%%%%%%%%%%%%%%%%%%%%%%%%%%%%%%%%%%%%%%%%%%%%%%%%%%%%%%%%%%%%%%%%%%%%%%%%%%%%%%%%%
%%%%%%%%%%%%%%%%%%%%%%%%%%%%%%%%%%%%%%%%%%%%%%%%%%%%%%%%%%%%%%%%%%%%%%%%%%%%%%%%%%%%%%%%%%%%%%%%%%%%%%%%%%%%%%%%%%%%%%%%%%%
%
\bibliographystyle{IEEEtran}

%
%
%
%%%%%%%%%%%%%%%%%%%%%%%%%%%%%%%%%%%%%%%%%%%%%%%%%%%%%%%%%%%%%%%%%%%%%%%%%%%%%%%%%%%%%%%%%%%%%%%%%%%%%%%%%%%%%%%%%%%%%%%%%%%
%%%%%%%%%%%%%%%%%%%%%%%%%%%%%%%%%%%%%%%%%%%%%%%%%%%%%%%%%%%%%%%%%%%%%%%%%%%%%%%%%%%%%%%%%%%%%%%%%%%%%%%%%%%%%%%%%%%%%%%%%%%
%	Biographies
%%%%%%%%%%%%%%%%%%%%%%%%%%%%%%%%%%%%%%%%%%%%%%%%%%%%%%%%%%%%%%%%%%%%%%%%%%%%%%%%%%%%%%%%%%%%%%%%%%%%%%%%%%%%%%%%%%%%%%%%%%%
%%%%%%%%%%%%%%%%%%%%%%%%%%%%%%%%%%%%%%%%%%%%%%%%%%%%%%%%%%%%%%%%%%%%%%%%%%%%%%%%%%%%%%%%%%%%%%%%%%%%%%%%%%%%%%%%%%%%%%%%%%%
%
\end{document}